\documentclass{article}

\usepackage{graphicx}                    
\usepackage{rotating}
\usepackage{color}                       

\chardef\us=`\_

\newcommand{\lya}{Ly$\rm \alpha$}

\usepackage{natbib}

\usepackage{authblk}  
\usepackage{orcidlink}  

\title{The EUV Late-Phase: Statistical Results from 15~Years of Solar Dynamics Observatory Observations}

\author[1]{Harry J. Greatorex\orcidlink{0000-0002-5302-0887}%
\thanks{Corresponding author: h.greatorex@qub.ac.uk}}

\author[1]{Aisling N. O'Hare\orcidlink{0009-0004-8025-9545}}

\author[1]{Susanna Bekker\orcidlink{0000-0001-5302-2543}}

\author[1]{Ryan J. Campbell\orcidlink{0000-0001-5699-2991}}

\author[2]{Daniel C. Keane\orcidlink{0009-0000-4610-6394}}

\author[1]{Ryan O. Milligan\orcidlink{0000-0001-5031-1892}}

\affil[1]{\small{
Astrophysics Research Centre, School of Mathematics and Physics,
Queen's University Belfast, Belfast, United Kingdom, BT7 1NN
}}

\affil[2]{\small{
School of Physics and Astronomy,
University of Glasgow, Glasgow, United Kingdom, G12 8QQ
}}

\begin{document}
\maketitle

\markboth{Greatorex et al.}{Solar Physics Example Article}
\begin{abstract}

Since its launch in 2010, the \textit{Solar Dynamics Observatory} (SDO) has provided continuous, high-cadence, multi-wavelength observations of the Sun, capturing thousands of solar flares and offering new insights into coronal dynamics. Among the discoveries enabled by SDO is the EUV late-phase (ELP), characterised by a secondary enhancement in warm coronal emission occurring tens of minutes after the main flare. While recent work has demonstrated the relevance of the ELP for space weather, the statistical behaviours and physical origins are not fully understood. Here, we present the most comprehensive statistical analysis of the ELP to date, based on 15-years of Fe~{\sc{xvi}} (335~\AA) observations from the \textit{Atmospheric Imaging Assembly} onboard SDO (SDO/AIA). From a sample of 5335 isolated flares between 2010 and 2025, we identify and validate 467 ELP events. The overall ELP occurrence rate was found to be 9~\%, with no significant dependence on the solar cycle and only a modest enhancement in the low-mid M-class range. The ELP typically exhibited an onset delay of 19~minutes, a peak-to-peak delay of 88~minutes, and a duration of 93~minutes. Strong correlations were found between ELP rise and decay rates ($\rho=0.76$), and between flare and ELP impulsivity ($\rho=0.61$). However, a comprehensive pairwise analysis revealed no significant correlation between the flare and ELP phases ($|\rho|<0.3$). A Principal Component Analysis of flare and ELP properties identified several semi-independent axes of variability, broadly associated with late-phase temporal scale, impulsive heating characteristics, and the relative prominence of flare and late-phase intensity measures. These results highlight the continuing importance of SDO’s long-term, high-resolution observations for uncovering new aspects of solar flare evolution and improving understanding of the Sun-Earth connection.

\end{abstract}

%


%

\section{Introduction} \label{sec:introduction}

Since its launch in February 2010, the Solar Dynamics Observatory (SDO; \citealt{Pesnell2012SDOMain}) has provided continuous, high-cadence imaging and spectroscopic coverage of extreme ultraviolet (EUV) emission from the Sun, revealing the complex dynamics of the solar atmosphere. In the standard model of solar flares (CSHKP; \citealt{Carmichael1964CSHKP, Sturrock1966CSHKP, Hiryama1974CSHKP, KoppPneuman1976CSHKP}), the reconnection of opposing polarity coronal magnetic fields releases stored magnetic free-energy, accelerating charged particles along newly formed closed loops, which deposit energy in the chromosphere through Coulomb collisions with dense plasma at the loop footpoints \citep{Aschwanden2002ElectronAcc, Holman2011Bremsstrahlung}. The resulting energy deposition drives rapid heating to temperatures exceeding $10^7$~K and the explosive expansion of chromospheric plasma into the overlying coronal loops with upflow velocities of several hundred km~s$^{-1}$ (Chromospheric Evaporation; \citealt{Fisher1985ChromosphericEvaporation, Milligan2006ChromosphericEvaporation, delZanna2006ChromosphericEvaporation, Milligan2009ChromosphericEvaporation, Tian2015ChromosphericEvaporation, Reep2016ChromosphericEvaporation}), which subsequently radiates in the soft X-ray (SXR) regime.

Flare evolution can be divided spectrally, spatially, and temporally into a series of phases. The preflare phase marks the accumulation of magnetic free-energy and is occasionally accompanied by hot onsets, which are instances of localised high-temperature emission preceding the impulsive rise \citep{Hudson1991HotOnsets, Hudson2021HotOnsets}. The impulsive phase corresponds to the rapid release of energy, producing intense nonthermal bremsstrahlung hard X-rays (HXRs) and chromospheric EUV emission emanating from the region surrounding the footpoints at the base of the flare loops. This is followed by the gradual phase, characterised by the dominance of SXR emission from hot flare loops. In accordance with the Neupert effect \citep{Neupert1968neuperteffect}, the impulsive phase HXR flux approximates the time derivative of the gradual phase SXR flux, with the observed delay between the two peaks reflecting the timescale of chromospheric evaporation. Temperatures within flaring loops can exceed 10--20~MK, in contrast to the ambient corona (1--3~MK). Subsequent loop cooling has been found to occur initially via thermal conduction, followed by radiative cooling as the temperature drops to EUV emitting temperatures \citep{Klimchuk2006CHP, Raftery2009CoronalCooling}. As cooling progresses, high-temperature EUV lines diminish and cooler lines become more prominent.

Extending beyond the gradual phase, the EUV late-phase (ELP) refers to a secondary peak of warm coronal emission, such as Fe~{\sc{xv}} and Fe~{\sc{xvi}}, occurring several minutes to a few hours after the initial solar flare \citep{Woods2011ELP, Woods2014EUVLatePhase}. The ELP is empirically defined by several key characteristics first outlined by \cite{Woods2011ELP}: (1) a delayed secondary peak in warm coronal EUV irradiance occurring tens of minutes to hours after the GOES SXR maximum; (2) the absence of a corresponding second enhancement in SXRs or in hot flare lines, such as Fe~{\sc{xx}} and Fe~{\sc{xxiii}}; (3) an association with an eruptive event such as a coronal mass ejection or coronal dimming; and (4) spatially distinct emission from a separate system of large, high-altitude loops that brighten well after the main flare arcade. 

The discovery of the ELP early in the SDO era raised fundamental questions surrounding the transport of energy in the solar atmosphere post-flare, yet relatively few studies have examined the physical origin of this emission. Two prevailing theories have been proposed. The first attributes ELP emission to the cooling of large, overlying coronal loops that were heated during the main phase of the flare, with their long cooling times reflecting the large spatial extent of these loops \citep{Hock2012ELP, Dai2018ELPHeatingCooling, Zhong2021ELP_AltMethods}. The second suggests that the ELP results from continued magnetic reconnection during the decay phase of the flare, with smaller energy releases generating renewed emission in warm EUV wavelengths \citep{Sun2013ELPReconnection, Dai2013ELP_Reconnection, Wang2026ELPSigmoid}. These two scenarios are not mutually exclusive, and evidence for both processes operating within the same active region has been observed \citep{Zhou2019ELPReconnection}. 

Not all flares possess an ELP; \cite{Woods2011ELP} found the phenomenon to occur in 14~\% of solar flares, with a greater probability for events of higher magnitude. A further study by \cite{Woods2014EUVLatePhase} found that ELP flares occur in a consistently higher proportion before and after solar maximum than during it, when using dual-decay events in SXRs as a proxy for ELP. More recently, a statistical analysis of $\sim$1800 disk-integrated flare observations from the \textit{Extreme-Ultraviolet Variability Experiment} (EVE; \citealt{Woods2012EVE}) instrument onboard SDO by \cite{Ornig2025ELPStats}, found that approximately 10~\% of flares exhibited a late-phase, and that most were classified as extreme ELP flares, where the flux of warm coronal emission in the late-phase exceeded that in the main phase \citep{Liu2015ExtremeELP}.

It is well established that both EUV and SXR emissions from solar flares have a measurable impact on the Earth's ionosphere. The altitude at which this radiation is effective depends on several factors, including the wavelength of the emission. Higher energy photons penetrate deeper into the atmosphere, so enhancements in SXR flux during solar flares primarily affect the lower ionosphere (D-region) and are routinely identified through perturbations in Very Low Frequency (VLF) radio signals \citep{Pant1993Dregion, McRae2005Dregion, Raulin2013LyaIonosphere, Hayes2021Ionosphere, Bekker2023Dregion}. At higher altitudes, within the E- and F-regions, increases in ionospheric electron density, which are predominantly driven by EUV flare emission at wavelengths shortward of 350~\AA\ \citep{Mitra1974Ionosphere, Leonovich2002Ionosphere, Zhang2011Ionosphere, Watanabe2021Ionosphere, Solomon2005Ionosphere}, are observed in total electron content (TEC) measurements derived from the Global Navigation Satellite System (GNSS). The Fe~{\sc{xv}} and Fe~{\sc{xvi}} lines that characterise the ELP are known to be geoeffective and therefore represent a key source of delayed ionospheric heating. Recent studies have found that ELP emission can produce a measurable enhancement in TEC (\citealt{Liu2024ELPIonosphere, Bekker2024ELP}). It has been demonstrated that the ionospheric response associated with the ELP can even exceed that of the main phase, particularly if the flare occurs near the solar limb, where significant attenuation of the cold chromospheric lines that ionise the ionosphere during the impulsive phase, is usually observed \citep{Bekker2025ELPCLV}. These findings highlight that the EUV late-phase can contribute significantly to the total flare energy deposited into the upper atmosphere. Understanding the temporal behaviour, occurrence frequency, and physical drivers of ELP emission is therefore essential for improving models of ionospheric variability and for advancing space weather forecasting capability.

Despite the apparent importance of the ELP for space weather, there remain relatively few studies of this phenomenon, which may partly reflect its inherent rarity. Since the discovery works by \citet{Woods2011ELP} and \citet{Woods2014EUVLatePhase}, SDO has captured thousands of additional flares. However, following the loss of the higher-energy MEGS-A channel of the \textit{Multiple EUV Grating Spectrograph} (MEGS) on the EVE instrument in 2014, continuous full-disk spectral measurements across the 50-370~\AA\ range were no longer available. While the MEGS-B channel coverage was subsequently extended down to approximately 330~\AA, EVE currently operates on a flare trigger system with a limited duty cycle. This trigger is initiated when the EUV flux measured by the \textit{Extreme Ultraviolet Spectrophotometer} (ESP; \citealt{Didkovsky2012ESP}) exceeds levels approximately equal to those associated with an M1.0 class flare. This flare-triggered mode will not capture smaller C-class flares and is not guaranteed to capture the full evolution or the extended timescales of the late-phase. Consequently, it is not feasible to perform a comprehensive long-term statistical study of ELP occurrence using SDO/EVE data alone. This limitation presents a unique opportunity to apply similar techniques to spatially-resolved imaging from the \textit{Atmospheric Imaging Assembly} (AIA; \citealt{Lemen2011AIA, Boerner2012AIA}), which has observed several of the key ELP wavelengths identified by \citet{Woods2011ELP} with continuous full-disk coverage since 2010.

Here we present a comprehensive analysis of more than 5000 isolated solar flares observed by SDO/AIA between 2010 and 2025, providing partial coverage of two solar cycles. This approach overcomes the limitations of previous studies, allowing us to examine the long-term behaviour of the ELP and to investigate potential solar cycle dependencies. The use of spatially-resolved measurements removes the ambiguity inherent in disk-integrated irradiance data, enabling us to directly characterise late-phase emission from individual flare regions. We also perform a comparative analysis between SDO/EVE and SDO/AIA detections from 2010--2014 to quantify the impact of instrument choice on measured ELP occurrence and to assess the extent to which disk-integrated observations may overestimate or misidentify late-phase events. Our approach focusses on the warm coronal Fe~{\sc{xvi}} (335~\AA) emission, which is known to be geoeffective \citep{Bekker2024ELP}. We derive observational metrics that describe the general behaviour of Fe~{\sc{xvi}} ELP emission, quantify its occurrence rate, class dependence, and solar cycle variation, and explore the relationships between main phase and late-phase emission characteristics. Particular attention is given to parameters that directly influence the ionosphere, such as impulsivity, onset delays, and decay times. Together, these analyses establish Fe~{\sc{xvi}} as a focal diagnostic of the EUV late-phase, providing a comprehensive, long-term characterisation of its behaviour. The results form a foundation for future investigations into the space weather impacts of late-phase emission and offer a framework for targeted analyses of the phenomenon. The observational data, event selection criteria,  processing methods, and procedures used for ELP detection, validation, and the derivation of temporal and morphological parameters are described in Section~\ref{sec:observations_methods}. The statistical results, including occurrence rates, solar cycle variation, temporal relationships, and impulsivity analysis, are presented in Section~\ref{sec:results}. A discussion of these findings in the context of previous work and potential space weather implications is given in Section~\ref{sec:discussion}. Finally, the main conclusions are summarised in Section~\ref{sec:conclusions}.

\section{Observations and Methods}\label{sec:observations_methods}

The following section outlines the data selection and analysis techniques used in this study, including the construction of the flare sample, processing of observation data, ELP identification methods, and the characteristic metrics calculated for validated ELP events.

\subsection{Flare Sample and Definition}\label{subsec:flare_sample}

To conduct a precise and comprehensive examination, we required a sample of isolated solar flares covering the entire operational lifetime of SDO to date. To be considered isolated, each flare was required to have no preceding events within 1~hour and no additional events for at least 3~hours after its end time, where these timings were defined by the SXR observations from the \textit{X-ray Sensor} (XRS; \citealt{Hanser1996GOES_XRS}) on board the primary \textit{Geostationary Environmental Operational Satellite} (GOES) spacecraft. This isolation criterion ensured that any detected ELP signature was unambiguous and that a suitable background period could be identified for subtraction and normalisation of EUV fluxes, allowing for accurate determination of ELP presence. In addition, each flare was required to be of magnitude C1.0 or greater (as determined by the peak flux in the 1--8~\AA\ GOES/XRS measurements) to aid the identification and differentiation of flare and ELP signatures. From the GOES flare list, approximately 5500 isolated events were identified between 2010--2025. Examination of the SXR timeseries for each GOES flare revealed a number of events unsuitable for analysis due to data dropouts, transmission gaps, or anomalies within the isolation window. Additional events were removed due to missing or corrupted data in the EUV channels. Excluding these cases yielded a final sample of 5335 flares for subsequent analysis.

\subsection{Data Acquisition and Processing}\label{subsec:data_acquisition}

Flare locations were obtained from the Heliophysics Event Knowledgebase (HEK; \citealt{Hurlburt2012HEK}) in helioprojective coordinates. For each flare, L~1 AIA image data were accessed through the \textit{Joint Science Operations Center} (JSOC) using 500~$\times$~500~arcsec cutouts centred on the flare coordinates. Each dataset covered a 6~hour window to capture both the impulsive and gradual evolution of the flare and to monitor any subsequent ELP activity. The data were sampled at 60~s cadence across four EUV wavelengths: 131~\AA, 171~\AA, 304~\AA, and 335~\AA. These wavelengths sample emissions across a broad range of plasma temperatures and mirror the methods of \cite{Woods2011ELP}. The 131~\AA\ channel is dominated by hot coronal lines (Fe~{\sc{xx}}, Fe~{\sc{xxiii}}), 171~\AA\ traces cooler coronal emission (Fe~{\sc{ix}}), 304~\AA\ samples the chromospheric and transition region response (He~{\sc{ii}}), and 335~\AA\ captures warm coronal emission (Fe~{\sc{xvi}}) that is present in the EUV late-phase. A summary of these channels is provided in Table~\ref{tab:aia_channels}.

\begin{table}[]
\footnotesize
\centering
\caption{Summary of the selected EUV channels from AIA, including the Full Width at Half Maximum (FWHM) of the imaging channel, the dominant emission species/constituent in that channel, the source region in the solar atmosphere, and the approximate emission temperature as a logarithm. Table adapted from \cite{Lemen2011AIA}.}
\begin{tabular}{ccccc}
\hline
\hline
\multicolumn{1}{c}{Channel (\AA)} & \multicolumn{1}{c}{FWHM ($\rm \Delta\AA$)} & \multicolumn{1}{c}{Constituent} & \multicolumn{1}{c}{Region} & \multicolumn{1}{c}{log(T)}\\
\hline
 131 & 4.4 & Fe~{\sc{xx}}, Fe~{\sc{xxiii}} & Flaring Corona & 7.0, 7.2\\
 171 & 4.7 & Fe~{\sc{ix}} & QS Corona and Upper-TR & 5.8\\
 304 & 12.7 & He~{\sc{ii}} & Chromosphere and TR & 4.7\\
 335 & 16.5 & Fe~{\sc{xvi}} & Coronal Active Regions & 6.4 \\
 \hline
\end{tabular}\label{tab:aia_channels}
\end{table}

For each event, the flux in each 60~s AIA cutout was integrated over the full field of view to yield a single total flux value per image, in Data Numbers (DNs), for each wavelength. These measurements were used to construct EUV flux timeseries spanning a 6~hour window centred on the GOES flare time. To isolate flare- and late-phase-related enhancements, a background subtraction was performed using a 1~hour preflare interval. This interval was chosen to ensure a sufficiently long and stable background baseline while avoiding contamination from earlier activity. A low-order polynomial (typically second or third order) was fit to this preflare segment to capture any slow-varying background trends, such as instrumental drift or gradual evolution in active region brightness. This fitted background was then extrapolated across the full 6~hour window and subtracted from the original timeseries. The result was then normalised to the preflare background level to express all flux enhancements as relative changes above quiescent conditions. To further reduce noise while preserving genuine flare-related structure, a Savitzky-Golay filter \citep{SavitzkyGolay1964} was applied to the timeseries using a third-order polynomial and a smoothing window of approximately 10~minutes. The resulting normalised fluxes were combined with the corresponding SXR timeseries from GOES to form the basis for ELP identification.

\subsection{ELP Detection and Validation}\label{subsec:elp_detection}

To identify the presence of ELPs in the flare sample, a manual inspection of the combined timeseries was performed, with a specific focus on the Fe~{\sc{xvi}} emission. A set of primary detection criteria were defined that were congruent with the first and second criteria outlined by \cite{Woods2011ELP} for SDO/EVE observations but adapted to fit the aims of this analysis. The criteria for a positive initial ELP detection were as follows:

\begin{enumerate}

    \item All positive ELP cases must have a distinct secondary peak in the 335~\AA\ channel flux that occurs after the initial flare event with no co-temporal increase in the SXR flux or the other EUV wavelengths.
    
    \item A clearly identifiable minimum point must exist in the 335~\AA\ flux between the flare end and ELP onset for all positive cases. However, we do not require the time delay between the flare end and ELP start to exceed a minimum threshold as this is a metric we intend to measure. 

    \item The magnitude of the secondary peak must be greater than 10~\% of the initial flare peak in 335~\AA. This threshold is lower than that adopted by \cite{Ornig2025ELPStats}; however, the typical short term variability of the background-subtracted Fe~{\sc{xvi}} flux in our smoothed AIA lightcurves is only a few percent, so a 10~\% enhancement represents a clear, statistically significant deviation above both the preflare background and instrumental noise. Given the space weather implications of ELP emission, where even small changes in EUV flux can have a detectable impact on the ionosphere \citep{Ohare2025QPPs}, this smaller threshold was considered sufficient.

\end{enumerate}

\begin{figure}[h!]
\centering
\includegraphics[width=1.\linewidth]{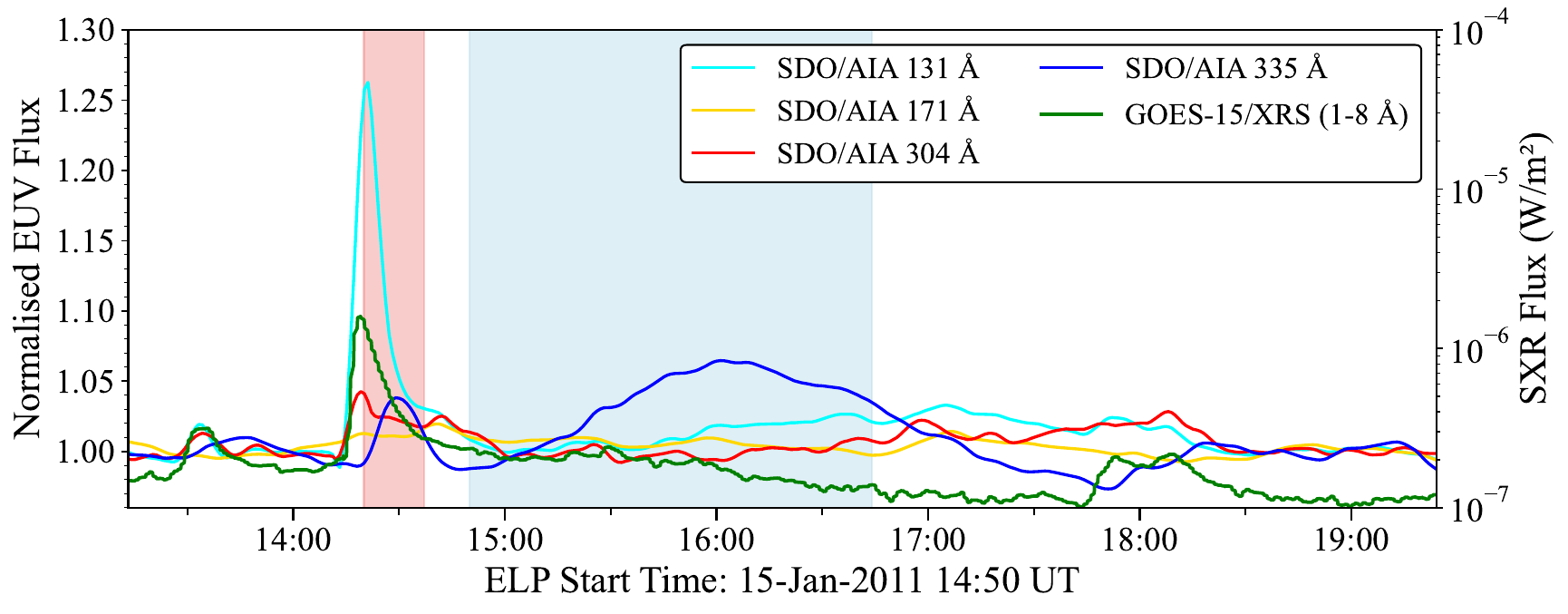}
\caption{Example of a combined EUV and SXR timeseries for a C1.1 flare corresponding to a positive ELP identification. Each EUV line was derived from integrated SDO/AIA cutout data. From the Fe~{\sc{xvi}} line (blue), two clear peaks are resolvable corresponding to the flare and ELP, respectively. The light red and light blue shaded areas represent the flare and ELP periods based on the definitions outlined in Section~\ref{subsec:elp_detection}, respectively. EUV fluxes were normalised to the preflare background period, which was determined in each wavelength from a low-order polynomial fitted to a 1~hour preflare period that was then extrapolated across the 6 hour window.}
\label{fig:elp_example}
\end{figure}

Consistent with other statistical investigations of the ELP \citep{Woods2014EUVLatePhase, Chen2020ELP, Ornig2025ELPStats}, the requirement that the flare be eruptive (criterion 3 of \citealt{Woods2011ELP}) was neglected, as this condition is not essential for identifying potentially geoeffective ELP emission and would introduce unnecessary complexity for a dataset of this scale. For each candidate, the flare and ELP start, peak, and end times were determined directly from the Fe~{\sc{xvi}} lightcurves. The flare start time was defined as the first sustained rise in Fe~{\sc{xvi}} exceeding a few percent above the preflare background for at least three consecutive 60~s bins. The flare peak time was defined as the point at which the initial Fe~{\sc{xvi}} increase reached its maximum value. The flare end time was defined as the point at which the initial Fe~{\sc{xvi}} emission decayed to approximately 50~\% of its maximum value. The ELP onset corresponded to the turning point following the flare decay where the Fe~{\sc{xvi}} emission began a renewed increase. Similar to the flare, the ELP peak time corresponded to the time of maximum Fe~{\sc{xvi}} flux thereafter, and the ELP end time corresponded to the point at which the Fe~{\sc{xvi}} flux had decayed to 50~\% of its late-phase peak value. While this requirement of a 50~\% decay is not necessarily physically prescriptive, it provides a consistent, reproducible metric across the full sample and allows direct comparison between both phases. An example of a combined EUV and SXR timeseries for a positive ELP identification is presented in Figure~\ref{fig:elp_example}. 

\begin{figure}[h!]
\centering
\includegraphics[width=1.\linewidth]{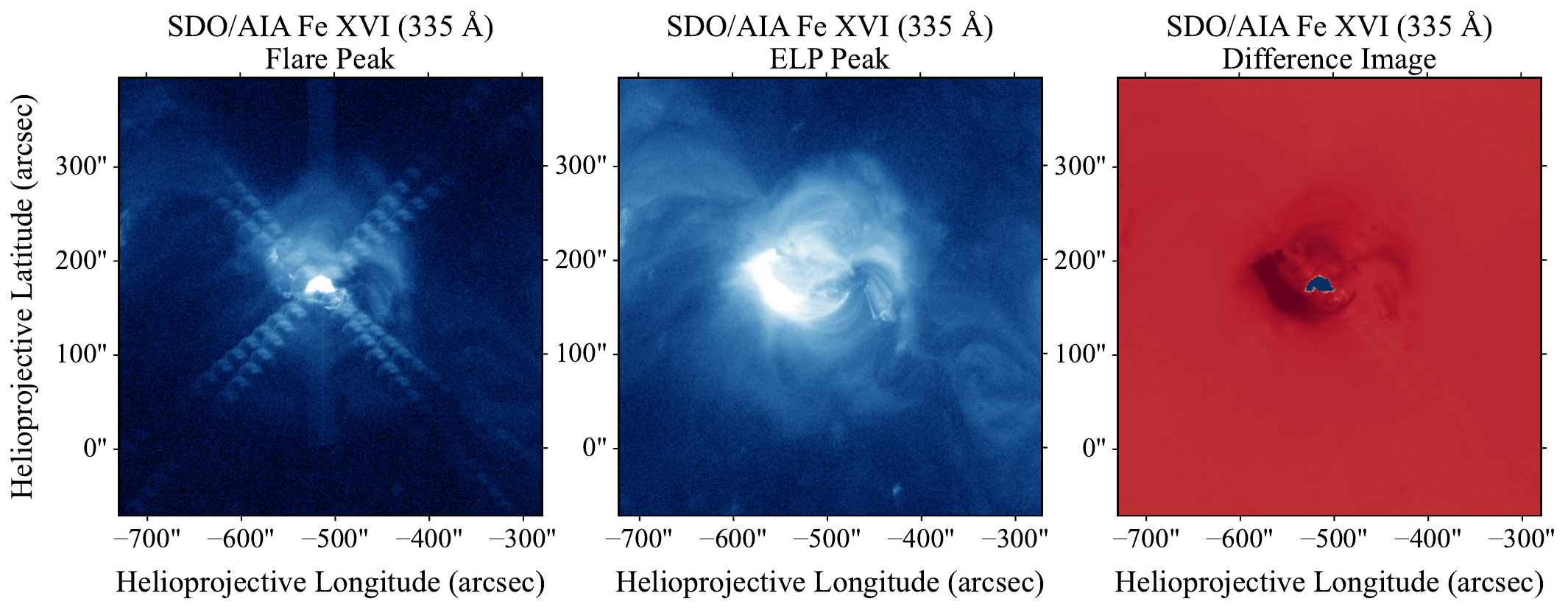}
\caption{Example of the spatial distinction between the main flare and the ELP emission in SDO/AIA Fe~{\sc{xvi}} (335~\AA) images. The left and middle panels show the 500~$\times$~500~arcsec AIA cutouts at the flare peak and the ELP peak, respectively. The right panel shows the difference image (ELP peak minus flare peak), where the dark region highlights the location of the enhanced late-phase emission, distinct from the brighter main phase region (blue).}
\label{fig:three_panel_validation}
\end{figure}

Flares meeting these conditions were labelled as ELP candidates. Each candidate was then subjected to a validation stage corresponding to the fourth criterion of \cite{Woods2011ELP}, requiring that the ELP emission originate from a spatially distinct loop system. AIA 335~\AA\ images at the flare and ELP peak times were compared manually, along with a difference image of the two. Valid ELP events exhibited clearly separated emission structures, with the late-phase loops overlying or adjacent to the main flare loop system. An example of two clearly distinct emission regions corresponding to the flare and ELP event in the same active region is presented in Figure~\ref{fig:three_panel_validation}. Here, the two brightenings originate from clearly separate loop systems, with the late-phase arcade appearing at higher altitude and displaced relative to the main flare loops. Cases lacking distinct spatial separation or displaying only diffuse post-flare brightening were rejected as misidentifications. In practice, these ambiguous detections typically occurred in weak events with marginal ELP flux enhancements or in large, fragmented active regions.

\subsection{Analysis of ELP Characteristics}\label{subsec:elp_metrics}

Following identification and validation of the ELP events, a series of quantitative parameters were calculated to describe their behaviour. These metrics were designed to capture the key physical properties of both the flare and late-phase, and to allow assessment of their potential space weather relevance. Event timings were defined with respect to the temporal evolution of the Fe~{\sc{xvi}} flux as observed in the 335~\AA\ channel of AIA, providing a consistent diagnostic of warm coronal emission independent of GOES SXR timings. The start, peak, and end times of both the flare and ELP were used to derive a set of temporal parameters. These included the onset delay (the interval between the end of the flare and the start of the ELP), the peak delay (the interval between the flare and ELP maxima), and the respective rise, decay, and total durations of each phase. To describe the temporal morphology of the ELP lightcurve, a skew parameter was defined as the ratio of the ELP rise time to the ELP decay time, where values greater than unity indicate a more impulsive rise and slower decay. To further characterise the late-phase evolution, rise and decay rates were also calculated from the Fe~{\sc xvi} lightcurves. The rise rate was defined as the flux-change per unit time between the ELP start and peak, and the decay rate as the corresponding flux-change between the ELP peak and end. 

In addition, the relative enhancement of Fe~{\sc{xvi}} emission for each phase was calculated as the ratio of the peak flux to the fitted preflare background ($\rm F_{peak}/F_{bg}$), providing a measure of the fractional brightening above the local background. The impulsivity, I, of each phase was calculated following the definition of \cite{Tamburri2024Impulsivity}, adapted here for Fe~{\sc{xvi}} emission as:

\begin{equation}
    \rm I = \frac{(F_{peak} - F_{start})/F_{bg}}{(t_{peak} - t_{start})}
\end{equation}

\noindent
where here, $\rm F_{start}$ is the Fe~{\sc{xvi}} flux as measured from SDO/AIA 335~\AA\ observations at the start time of the respective phase, $\rm t_{start}$, as defined by the criteria outlined in Section~\ref{subsec:elp_detection}. Similarly, $\rm F_{peak}$ is the peak Fe~{\sc{xvi}} flux in each respective phase occurring at time $\rm t_{peak}$, and $\rm F_{bg}$ is the background flux calculated from the preflare period defined in Section~\ref{subsec:elp_detection}. The global averages reported in this study are accompanied by the standard deviation ($\sigma$) of each parameter across the full ELP sample. These values reflect the intrinsic variability of the population rather than representing formal measurement uncertainties. Timing precision is constrained to the selected 60~s cadence of the AIA data, introducing an uncertainty of approximately $\pm$30~s in start and peak times. Flux-related metrics inherit this temporal uncertainty and an additional few percent variability from instrumental noise and background normalisation. These uncertainties are small compared with the overall spread of the measured distributions. To estimate the impact of AIA saturation on the flux-related measurements, the DESAT reconstruction method of \citet{Schwartz2015desat} was applied to a small test sample of ten saturated flares. For each event, the enhancements derived from the original saturated and reconstructed images were compared. The enhancement difference, $\Delta E$, ranged from 0.02 to 0.14, with a mean values of 0.07. This additional uncertainty is applied to the calculated flux-related metrics. It should be noted that most flares in the sample did not suffer from saturation effects, particularly not during the late-phase, where the emission region is typically larger and more diffuse. Ultimately, these systematic uncertainties are included for completeness and to exercise precautionary estimates, but they do not impact the statistical trends discussed in the following section.

To test for relationships among the timing and morphological parameters, a pairwise correlation analysis was performed on all validated ELP events. The variables included the flare and ELP durations, onset and peak delays, ELP rise and decay times, and the skew parameter. For each parameter pair, the Spearman ($\rm \rho$; \citealt{Spearman1904}) correlation coefficients were computed, with uncertainties estimated through a 1000-sample bootstrap procedure that provided $1\sigma$ confidence intervals for each coefficient. The resulting correlation matrices were examined to identify statistically significant and physically meaningful trends among the derived parameters. To explore broader structure among the measured flare and ELP parameters, a principal component analysis (PCA; \citealt{Jolliffe2011PCA}) was also performed on the events with confirmed ELP. Thirteen variables, derived from the above analysis steps, spanning timescales, impulsivities, radiative enhancements, and rates were included. Each variable was standardised prior to decomposition to prevent scale-driven bias. The first three principal components were then retained and used to assess variability across the sample.

\section{Results}\label{sec:results}

The following section presents the results of a statistical analysis of the validated ELP events. We first describe their overall occurrence and dependence on flare magnitude and solar cycle phase, before comparing detection statistics between SDO/AIA and SDO/EVE. We then examine the temporal, morphological, and flux-related characteristics of the ELP derived from the Fe~{\sc{xvi}} metrics defined in Section~\ref{sec:observations_methods}.

\subsection{ELP Occurrence Rates}\label{subsec:occurence}

From examination of the AIA timeseries for the 5335 total flares in the sample, 515 events (9.65~\%) were initially classified as potential ELP flares. Following a manual examination of the associated 335~\AA\ images, 467 events were confirmed as ELPs (8.75~\%). In terms of flare magnitude, ELPs were detected in 6.99~\% of C-class, 19.32~\% of M-class, and 19.18~\% of X-class flares, indicating a preference towards moderate-to-large events. However, the proportions for higher magnitude flares are subject to larger statistical uncertainty due to their smaller sample sizes. Table~\ref{tab:elp_occurrence} summarises the detection statistics for each GOES class.

\begin{table}[h!]
\centering
\caption{Summary of ELP occurrence rates by GOES class. The number of total flares ($\rm N$), detected ELPs ($\mathrm{ELP}_{\mathrm{Potential}}$), and validated ELPs ($\mathrm{ELP}_{\mathrm{True}}$) are listed, together with the corresponding percentages.}
\label{tab:elp_occurrence}
\begin{tabular}{lcccc}
\hline
\hline
GOES Class & $\rm N$ & $\mathrm{ELP}_{\mathrm{Potential}}$ & $\mathrm{ELP}_{\mathrm{True}}$ & $\rm ELP_{\%}$ \\
\hline
C & 4580 & 361 & 320 & 6.99 \\
M & 678 & 138 & 131 & 19.32 \\
X & 73 & 14 & 14 & 19.18 \\
\hline
Overall & 5335 & 515 & 467 & 8.75 \\
\hline
\end{tabular}
\end{table}

\begin{figure}[h!]
\centering
\includegraphics[width=1.\linewidth]{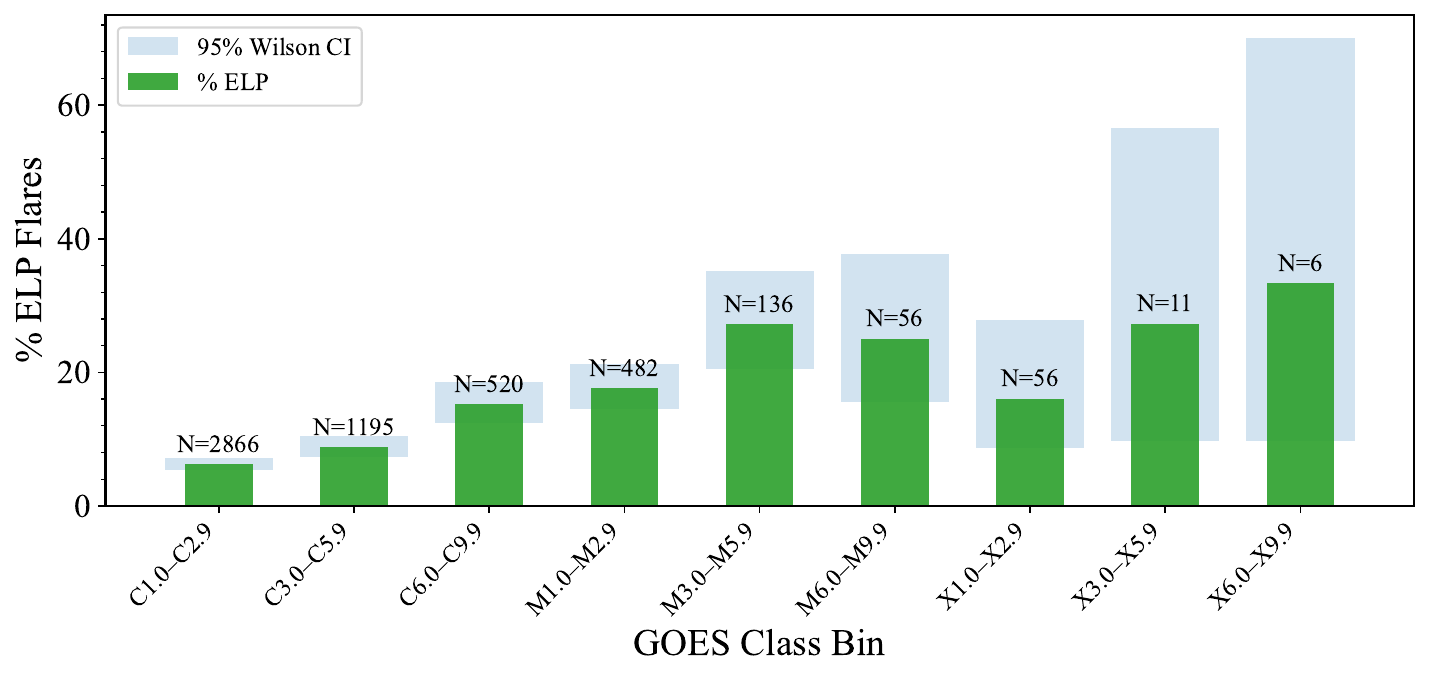}
\caption{Percentage of events with an associated ELP by GOES class bin. Error bars denote 95~\% Wilson binomial confidence intervals. The annotated N-values correspond to the total number of flares from the full sample of 5335 flares.}
\label{fig:elp_occurrence_bins}
\end{figure}

To further assess the class dependence of ELP emission, GOES classes were subdivided into finer class bins. Figure~\ref{fig:elp_occurrence_bins} shows the ELP fraction as a function of flare magnitude. The occurrence rate increased from 6.2~\% for C1.0--C2.9 events to a maximum of 27.2~\% for M3.0--M5.9 events before decreasing at higher magnitudes. A secondary rise is observed among low-to-moderate X-class flares; however, the reduced number of events in these categories introduces substantial statistical uncertainty. The error bars in Figure~\ref{fig:elp_occurrence_bins} represent 95~\% Wilson binomial confidence intervals, which provide a more accurate measure of proportional uncertainty for finite sample sizes than the standard normal approximation \citep{Wilson01061927}.

\begin{figure}[h!]
\centering
\includegraphics[width=1.\linewidth]{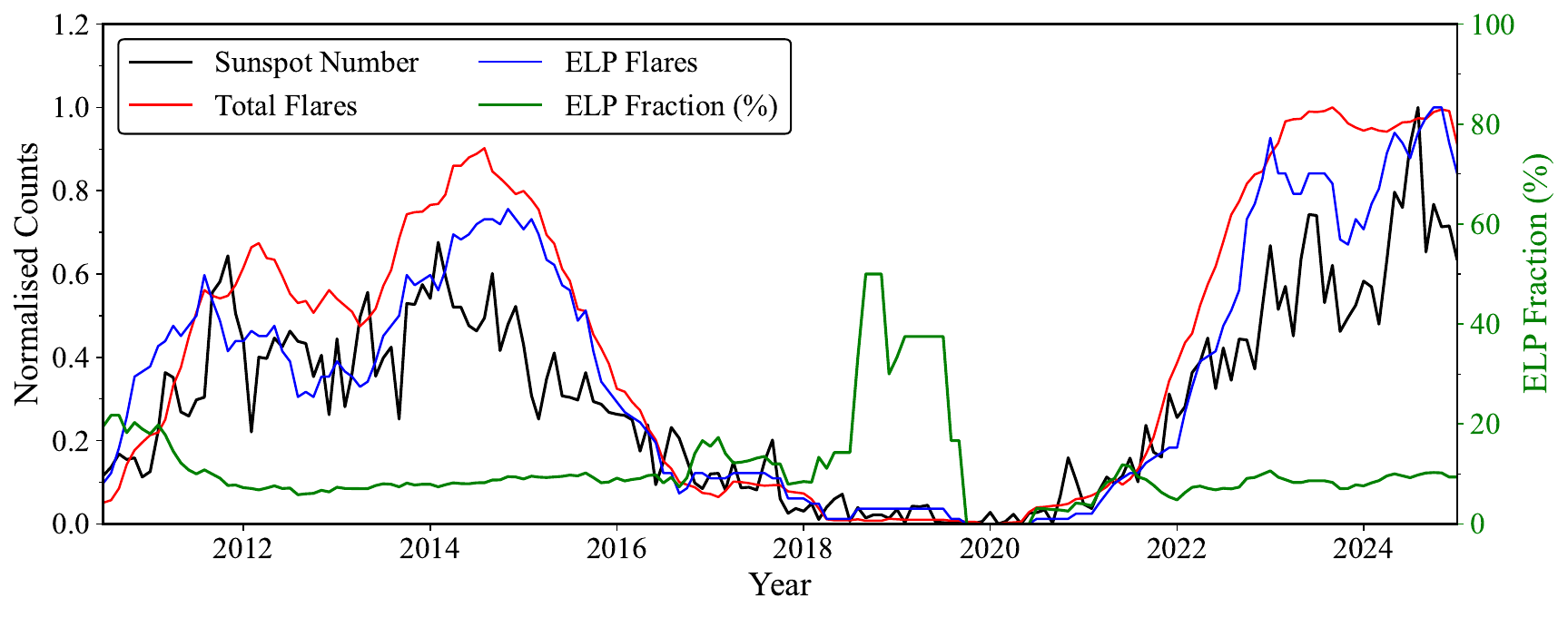}
\caption{Temporal variation of ELP occurrence from 2010 to 2025. The black line indicates the six-month smoothed monthly mean sunspot number used as a proxy for solar activity level and therefore solar cycle. The red line denotes the total number of flares per month, and the blue line shows the number of validated ELP detections in that month. The percentage of ELPs detected in each month (normalised to the total number of flares) is shown in green.}
\label{fig:elp_solarcycle}
\end{figure}

To investigate the long-term variability of ELP occurrence, the validated ELP sample was examined across the 15~year period covered by the dataset. The monthly mean sunspot number was used as a proxy for global solar activity and to define the phase of the solar cycle. The temporal variation in ELP occurrence was evaluated by computing the total number of validated ELP detections per month, the total number of flares observed, and the fraction of these flares exhibiting an ELP. Figure~\ref{fig:elp_solarcycle} presents the comparative evolution of solar activity and ELP occurrence, smoothed using a six-month running mean to emphasise large-scale trends and minimise short-term fluctuations.

Figure~\ref{fig:elp_solarcycle} demonstrates that there is no strong solar cycle dependence in the occurrence of ELPs. The solar cycle trend is clearly evident in the sunspot number, with activity levels rising and falling periodically and an approximate 11~year interval between successive peaks. The total flare count and the number of ELP detections track the changing global activity level, as indicated by the coherence between the black, red, and blue curves. The fraction of ELP events remains relatively uniform across the 15-year period, with an average value of $\sim$10~\%, consistent with the overall occurrence rate reported above. A local enhancement in the ELP fraction between 2018 and 2020 coincides with solar minimum; however, this feature occurs during a period of low flare numbers (47 isolated flares) and is therefore subject to statistical uncertainty. A larger flare sample during this epoch would be required to confirm whether this increase is significant.

\subsection{Cross-comparison of SDO/AIA and SDO/EVE}\label{subsec:AIA_EVE_comp}

To assess the instrumental impact on ELP identification, a comparison was performed between detections from SDO/AIA and SDO/EVE over the period of 2010--2014. This window covers the period during which the MEGS-A component of EVE was fully operational. These measurements are required here as the MEGS-A spectral range encapsulates all wavelengths examined for the ELP identification from AIA, allowing for a direct comparison of ELP detection capabilities using each instrument. All flares from the AIA sample with complete EVE coverage during this period were identified, yielding more than 500 events. A representative subsample of 500 flares was selected for analysis, comprising of approximately 86~\% C-class, 12~\% M-class, and 2~\% X-class events, thus maintaining a proportional class representation while reducing the computational demand of the comparison.

For the EVE analysis, Level~2B spectral line data at 60~s cadence were used to construct lightcurves with identical wavelengths to those in the AIA analysis and analysed over the same six hour windows. Background subtraction, normalisation, and smoothing were applied identically to those used for the AIA data to ensure consistency between the two datasets (see Section~\ref{subsec:elp_detection}). The same ELP detection criteria described in Section~\ref{subsec:elp_detection} were then applied to the EVE lightcurves.

\begin{figure}[h!]
\centering
\includegraphics[width=1.\linewidth]{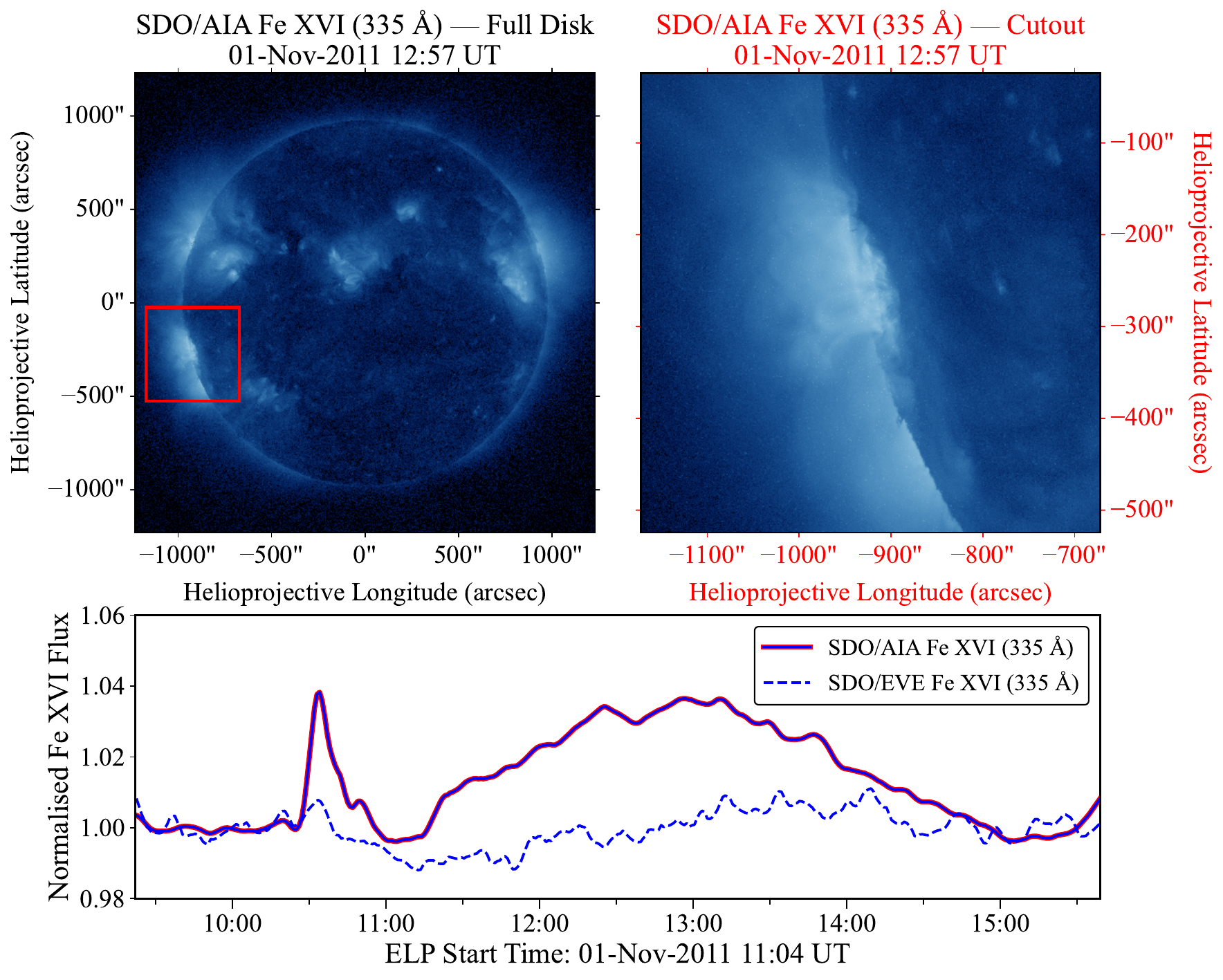}
\caption{Comparison between SDO/AIA and SDO/EVE observations of the Fe~{\sc{xvi}} emission for a representative ELP event. Top Left: Full-disk AIA 335~\AA\ image at the ELP peak time. The red box denotes the 500~$\times$~500~arcsec cutout region corresponding to the ELP/flare site used as a proxy for the effective field of view of the disk-integrated EVE measurements. Top right: AIA 335~\AA\ cutout image of the region enclosed by the red box in the full-disk image. Bottom: Normalised Fe~{\sc{xvi}} lightcurves from AIA (blue-red line) and EVE (blue dashed line) for the same event. The ELP signal is clearly visible in the spatially-resolved AIA data but not in the disk-integrated EVE irradiance.}
\label{fig:aia_eve_comp}
\end{figure}

Across the 500 flares examined, the classifications derived from EVE and AIA agreed in approximately 80~\% of cases. This agreement demonstrates broad consistency between the instruments but also highlights the potential discrepancies presented by using different measurements to identify ELPs. The spatially integrated nature of EVE means that emission from multiple active regions contributes to the total irradiance, which can obscure the Fe~{\sc{xvi}} emission associated with the ELP. Additionally, the associated background flux for disk-integrated measurements may be elevated by multiple active regions, potentially masking smaller ELP events. These effects lead to potential misidentification of the ELP, particularly during periods of high solar activity. Figure~\ref{fig:aia_eve_comp} exemplifies this, with the upper-left panel demonstrating a full-disk AIA image in 335~\AA\ of the Sun at the point of maximum ELP emission associated with a flare. This image represents the effective field-of-view (FOV) for the EVE instruments disk-integrated measurements. The image shows several active regions present on the disk and limb, with the red bounding box representing the subsequent AIA cutout region centred around the ELP active region analogous to those used for ELP identification in this study. The upper-right panel contains a 500~$\times$~500~arcsec AIA cutout image centered around the flare location. The bottom panel compares the AIA and corresponding EVE timeseries, where the dashed line corresponds to the disk-integrated irradiance from EVE, and the solid line denotes the derived timeseries from the AIA cutouts, each normalised to their respective background as described in Section~\ref{subsec:elp_detection}. From these plots, it is clear that  the enhancement in the EVE signal is diminished compared to that from AIA and thus no clear ELP can be detected from these observations. On the other hand, a clear ELP signal is identifiable in the spatially constrained AIA data. 

\begin{figure}[h!]
\centering
\includegraphics[width=1.\linewidth]{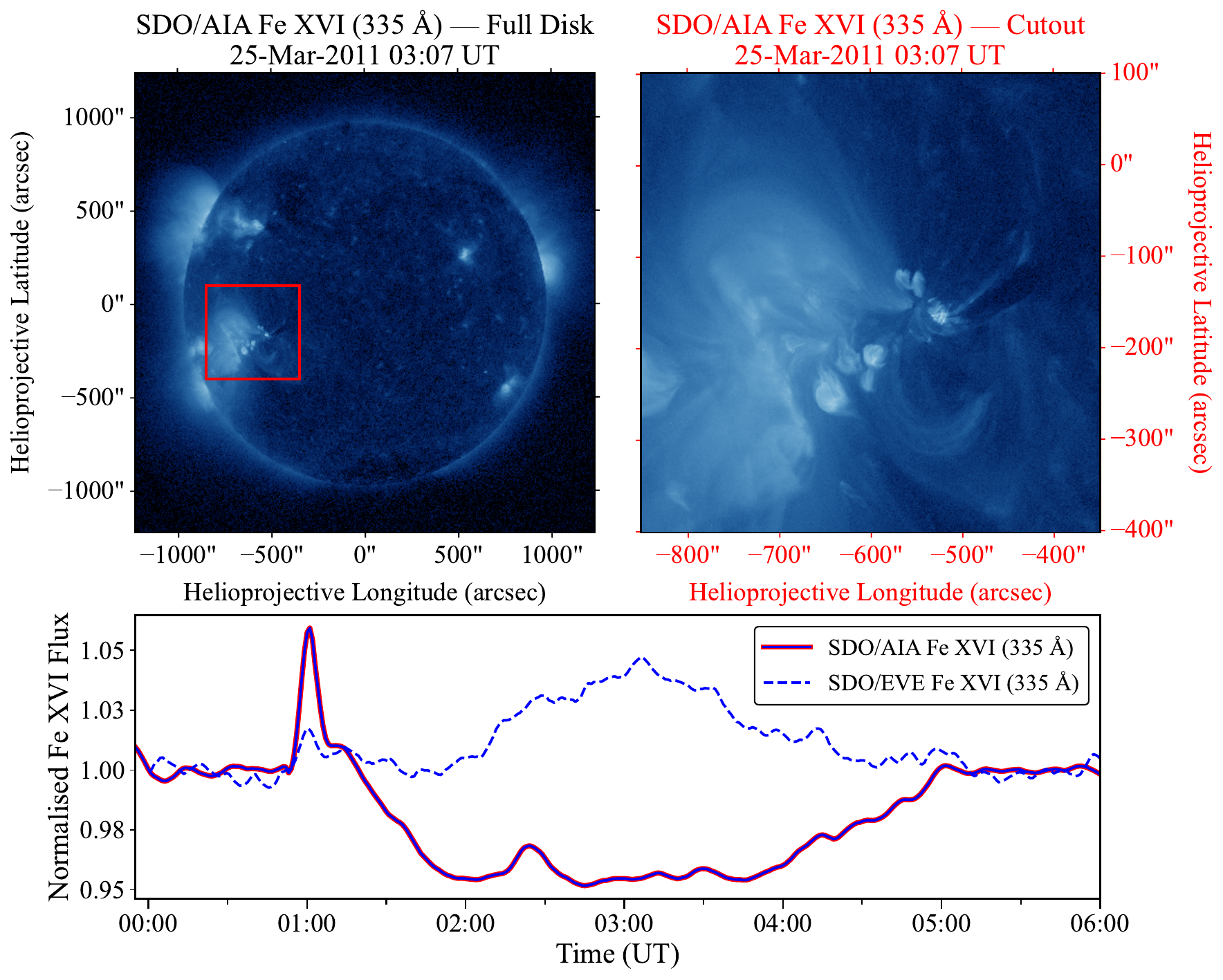}
\caption{Comparison between SDO/AIA and SDO/EVE observations of the Fe~{\sc{xvi}} emission for a representative ELP event. Top Left: Full-disk AIA 335~\AA\ image at the ELP peak time. The red box denotes the 500~$\times$~500~arcsec cutout region corresponding to the ELP/flare site used as a proxy for the effective field of view of the disk-integrated EVE measurements. Top right: AIA 335~\AA\ cutout image of the region enclosed by the red box in the full-disk image. Bottom: Normalised Fe~{\sc{xvi}} lightcurves from AIA (blue-red line) and EVE (blue dashed line) for the same event. An ELP signal is visible in the disk-integrated EVE irradiance but the spatially-resolved AIA data demonstrates a dip with no signal, potentially associated with an absence of emitting material due to the initial flare eruption.}
\label{fig:aia_eve_comp_2}
\end{figure}

While the example shown in Figure~\ref{fig:aia_eve_comp} illustrates an ELP identified in AIA but not EVE, the inverse case also occurred in a small subset of events. These instances likely arise when the AIA cutout FOV does not fully capture the spatial extent of the ELP emission, particularly for eruptive flares involving large-scale displacements or for active regions dispersed across the disk (Figure~\ref{fig:aia_eve_comp_2}). Additionally, coincident emission from multiple regions contributing to the EVE signal, but outside the AIA bounding box, may produce apparent late-phase signatures in EVE that are not visible in the AIA FOV. These alternative scenarios are discussed further in Section~\ref{sec:discussion}. For now, we affirm that the AIA results provide a more robust estimate of ELP occurrence.

\subsection{Temporal Properties of ELP Emission}\label{subsec:temporal_elp}

To characterise the evolution of the ELP, all timing analyses were performed using the Fe~{\sc{xvi}} lightcurves from AIA. The parameters examined here are related to the relative timing and duration of the ELP emission with respect to the main flare exclusively in Fe~{\sc{xvi}}. Table~\ref{tab:temporal_summary} contains a summary of the derived parameters, which are further discussed in Sections~\ref{subsubsec:onset_delay}--\ref{subsubsec:temporal_morph}.

\begin{table}[h!]
\footnotesize
\centering
\caption{Summary of temporal parameters derived from the observed Fe~{\sc{xvi}} emission for the 467 validated ELP events examined in this study. Uncertainties represent the standard deviation ($\sigma$) of each quantity across the sample. Each parameters is analysed in detail in their related subsections.}
\label{tab:temporal_summary}
\begin{tabular}{lcccc}
\hline
\hline
\multicolumn{1}{c}{Parameter} & \multicolumn{1}{c}{Mean $\pm$ $\sigma$} & \multicolumn{1}{c}{Median $\pm$ $\sigma$} & \multicolumn{1}{c}{Min} & \multicolumn{1}{c}{Max} \\
\hline
Onset Delay (min) & 19 $\pm$ 18 & 13 $\pm$ 5 & 2 & 119 \\
Peak-to-Peak Delay (min) & 88 $\pm$ 41 & 82 $\pm$ 10 & 8 & 226 \\
Flare Duration (min) & 23 $\pm$ 11 & 20 $\pm$ 6 & 4 & 61 \\
ELP Duration (min) & 93 $\pm$ 42 & 89 $\pm$ 15 & 15 & 239 \\
ELP Rise Time (min) & 61 $\pm$ 32 & 55 $\pm$ 10 & 8 & 180 \\
ELP Decay Time (min) & 31 $\pm$ 17 & 28 $\pm$ 8 & 4 & 148 \\
ELP Rise Fraction & 0.65 $\pm$ 0.15 & 0.66 $\pm$ 0.12 & 0.25 & 0.91 \\
ELP Decay Fraction & 0.35 $\pm$ 0.15 & 0.34 $\pm$ 0.12 & 0.09 & 0.75 \\
Skew & 2.36 $\pm$ 1.45 & 2.10 $\pm$ 0.9 & 0.15 & 9.00 \\
ELP-Flare Duration Ratio & 5.63 $\pm$ 2.3 & 4.95 $\pm$ 1.8 & 1.0 & 14.5 \\
\hline
\end{tabular}
\end{table}

\subsubsection{ELP Onset Delay}\label{subsubsec:onset_delay}

\begin{figure}[h!]
\centering
\includegraphics[width=0.8\linewidth]{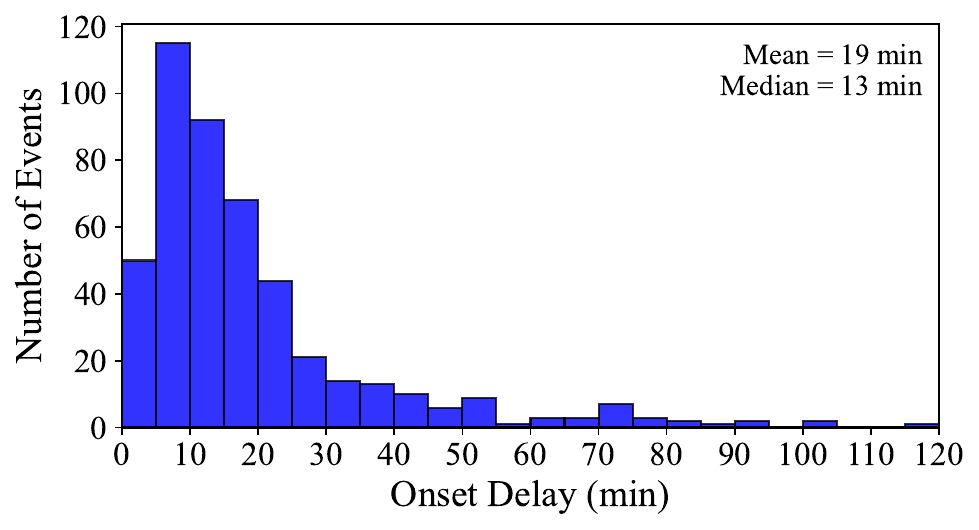}
\caption{Histogram of the ELP onset delay ($\mathrm{\Delta t_{\mathrm{onset}}}$) distribution in Fe~{\sc{xvi}}, constructed in 5~minute bins for the 467 validated ELP events. Mean and median values are displayed in the upper-right corner.}
\label{fig:onset_delay}
\end{figure}

Here, the onset delay ($\rm \Delta~t_{onset}$) refers to the duration between the end of the Fe~{\sc{xvi}} emission from the main flare and the start of the Fe~{\sc{xvi}} emission associated with the corresponding ELP. Figure~\ref{fig:onset_delay} presents a histogram of onset delay times for the 467 validated ELP events in the sample. The mean onset delay is 19~$\pm$~18~minutes, with a median value of 13~minutes. The large spread reflects the intrinsic variability of this parameter, with most events exhibiting short delays of less than 20~minutes, while a smaller number show significantly longer separations. Beyond $\rm \Delta~t_{onset}$~$>$~30~minutes, the frequency decreases rapidly, with fewer than 20 events in each successive 10~minute bin. The onset delay between flare-related and ELP-related Fe~{\sc{xvi}} emission is highly variable but typically remains within the first 20~minutes following the main phase of the flare.

\subsubsection{Peak-to-Peak Delay}\label{subssubsec:peak2peak}

\begin{figure}[h!]
\centering
\includegraphics[width=0.8\linewidth]{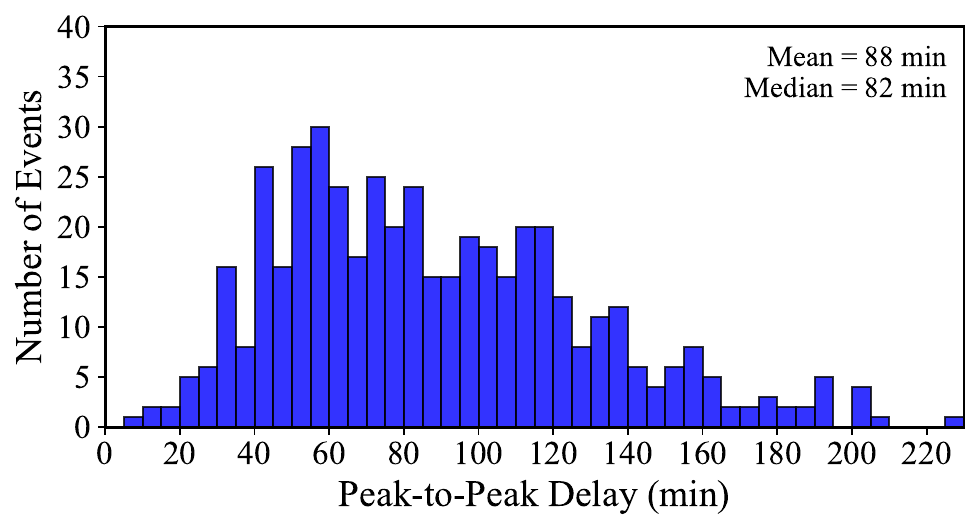}
\caption{Histogram of the peak-to-peak delay ($\mathrm{\Delta t_{{peak}}}$) between the Fe~{\sc{xvi}} emission maxima of the main flare and the corresponding ELP, constructed in 10~minute bins for the 467 validated events. Mean and median values are displayed in the upper-right corner.}
\label{fig:peak_delay}
\end{figure}

The peak-to-peak delay ($\rm \Delta t_{{peak}}$) represents the time interval between the maximum of the Fe~{\sc{xvi}} emission during the main flare and the maximum of the corresponding late-phase emission.
From Figure~\ref{fig:peak_delay}, it is apparent that the peak delays span a broad range from 8 to 226~minutes. A mean $\rm\Delta t_{{peak}}$ of 88~$\pm$~41~minutes and a median of 82~minutes were calculated. The distribution roughly resembles a normal distribution with a moderate asymmetry, with the majority of ELP peaks occurring within two hours of the flare maximum. Only a small number of events ($\sim3$~\%) exhibit peak separations greater than 180~minutes. 

\subsubsection{Flare and ELP Durations}\label{subsubsec:durations}

\begin{figure}[h!]
\centering
\includegraphics[width=0.85\linewidth]{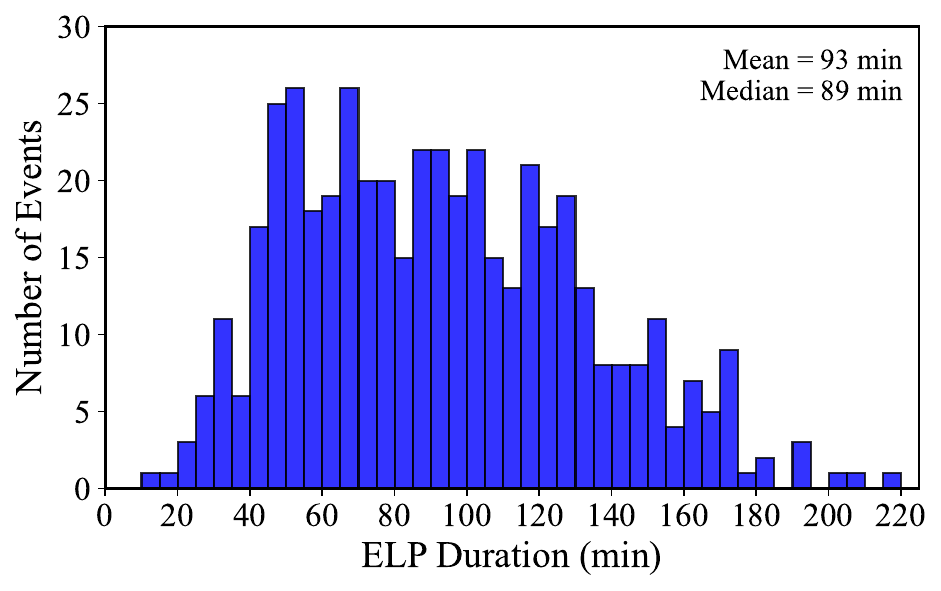}
\caption{Histogram of the ELP duration ($\mathrm{t_{ELP}}$) in Fe~{\sc{xvi}}, constructed in 5~minute bins for the 467 validated events. Mean and median values are displayed in the upper-right corner.}
\label{fig:duration_bar}
\end{figure}

\begin{figure}[h]
\centering
\includegraphics[width=0.7\linewidth]{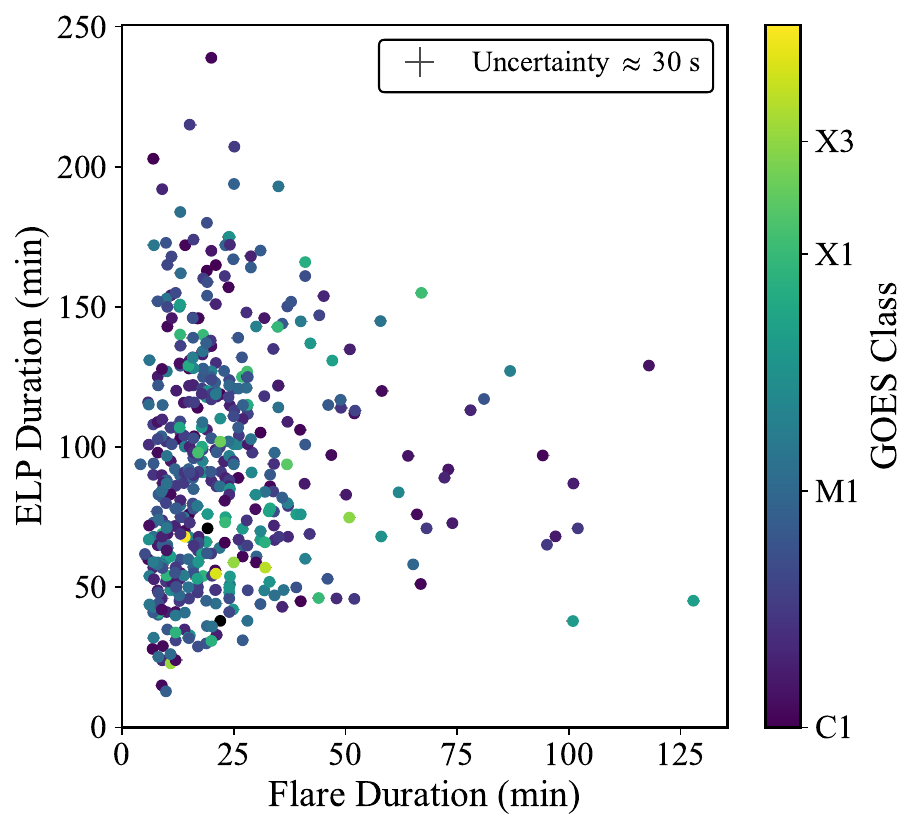}
\caption{Comparison between flare and ELP durations derived from Fe~{\sc{xvi}} emission for the 467 validated events. Each point represents an individual flare-ELP pair, coloured by GOES class. Individual measurement uncertainties arise from the 60~s AIA cadence (timing precision of $\pm$30~s), but these duration errors are small compared to the measurement values and therefore not visible at the scale of the plot.}
\label{fig:durations_scatter}
\end{figure}

Figure~\ref{fig:duration_bar} presents the distribution of the overall ELP duration ($\rm t_{ELP}$), defined as the interval between the start and end times of the secondary Fe~{\sc{xvi}} emission, following the definitions outlined in Section~\ref{subsec:elp_detection}. Across the sample, ELP durations range from 15 to 239~minutes, with a mean of 93~$\pm$~42~minutes and a median of 89~minutes. The distribution is approximately normal, with most events lasting between one and two hours. The flare durations were similarly defined from the Fe~{\sc{xvi}} emission, rather than from the standard GOES SXR timings, as the interval between the start and end of the primary Fe~{\sc{xvi}} brightening. In contrast to the ELP, flare durations are typically much shorter, with a mean of 23~minutes. The ELP duration was found to exceed the flare duration in 98~\% of events. The average duration ratio between the ELP and the associated flare ($\overline{R_t}$) was found by:

\begin{equation}
    \overline{R_t} = \frac{1}{N} \sum_{i=1}^{N} \frac{\rm t_{ELP, i}}{\rm t_{flare, i}} = 5.63 
\end{equation}

\noindent
where N is the number of valid ELP events, and $\rm t_{ELP, i}$ and $\rm t_{flare, i}$ are the ELP and flare durations per event, respectively. This indicates that, on average, the Fe~{\sc{xvi}} emission during the ELP persists for more than five times the duration of the main flare. Figure~\ref{fig:durations_scatter} compares the flare and ELP durations, with points coloured by GOES class, represented on a logarithmic scale for clarity. Individual 1$\sigma$ uncertainties in duration were estimated via 1000 Monte Carlo resamplings, perturbing each start and end time by Gaussian noise ($\sigma$ = 1~min). This value represents the approximate timing precision of the AIA derived lightcurve boundaries. The resulting per-event uncertainties are small compared to the total durations of tens to hundreds of minutes and are therefore not easily visible in Figure~\ref{fig:durations_scatter}. From this distribution, it is evident that the durations of both the flares and their corresponding ELPs are largely independent of flare magnitude and no clear correlation exists between the durations of the two phases.

\subsubsection{Temporal Morphology of the ELP}\label{subsubsec:temporal_morph}

\begin{figure}[h!]
\centering
\includegraphics[width=1.\linewidth]{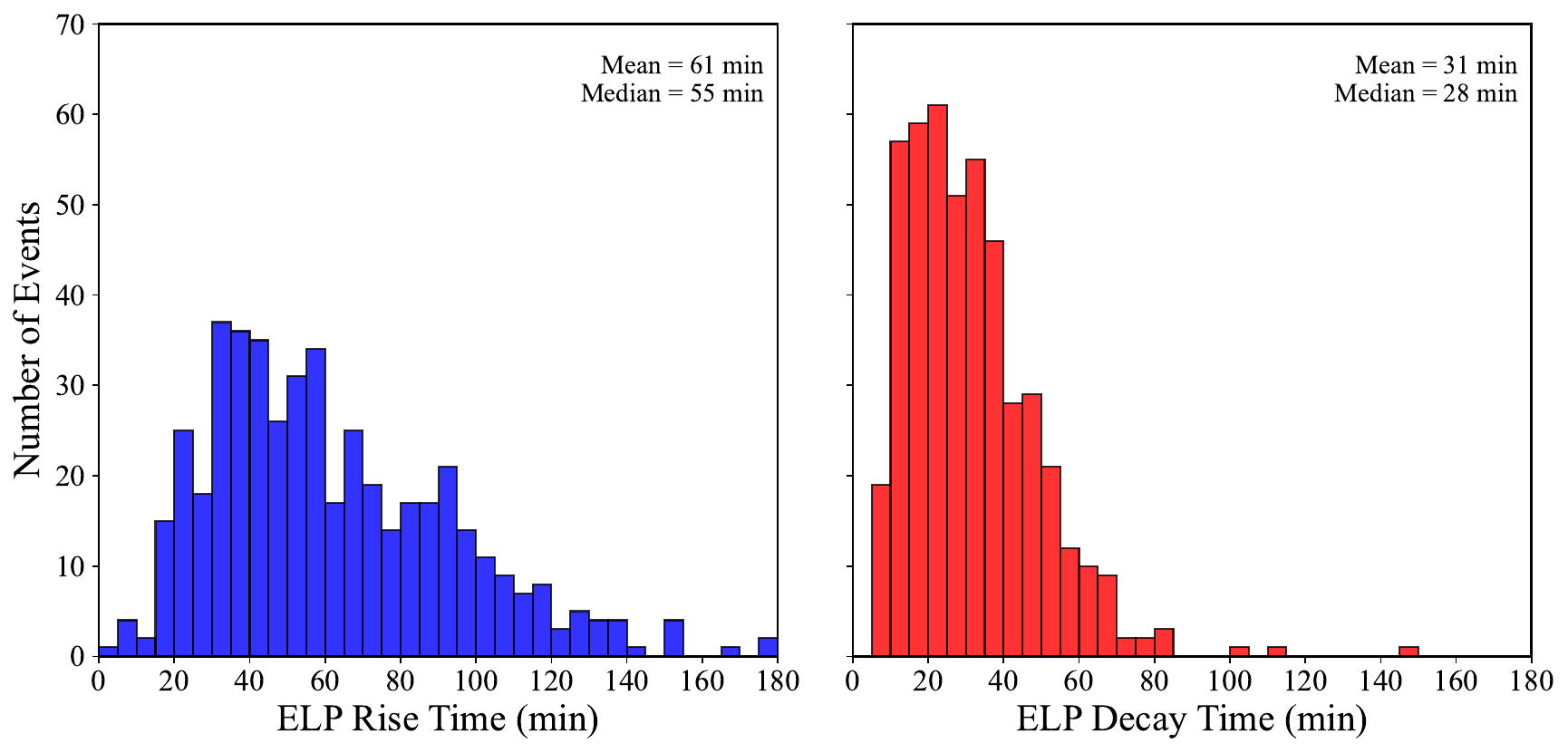}
\caption{Distributions of the Fe~{\sc{xvi}} late-phase rise time (left; blue) and decay time (right; red) for the 467 validated events. Each panel shows the corresponding histogram for that interval each in 5~minute bins. Mean and median values are displayed in the upper-right corner of each panel.}
\label{fig:rise_decay_time}
\end{figure}

The temporal evolution of the ELP provides important context for understanding the physical processes driving the emission, and several details can be inferred from the morphology of the Fe~{\sc{xvi}} timeseries. Figure~\ref{fig:rise_decay_time} illustrates the rise (left panel) and decay (right panel) times of the Fe~{\sc{xvi}} emission during the late-phase. Across the validated sample, the mean rise time was 61~$\pm$~32~minutes, with a median of 55~minutes, whereas the decay time was typically shorter, with a mean of 31~$\pm$~17~minutes and a median of 28~minutes. The corresponding ELP skew parameter, defined as $\rm t_{\mathrm{rise}}/\rm t_{\mathrm{decay}}$ (Section~\ref{subsec:elp_metrics}), has a mean value of 2.36, indicating that the rise phase is typically much longer than the decay phase. We note that the decay time is measured until the Fe~{\sc{xvi}} flux returns to 50~\% of its peak value, which may slightly underestimate the true duration of the decay phase, particularly in long-lived or slowly cooling events. This approach could introduce a systematic truncation that biases the skew and duration statistics modestly toward shorter values. Nevertheless, this threshold ensures consistency across the sample and mitigates the influence of background drift or post-flare variability.

\begin{figure}[h!]
\centering
\includegraphics[width=1.\linewidth]{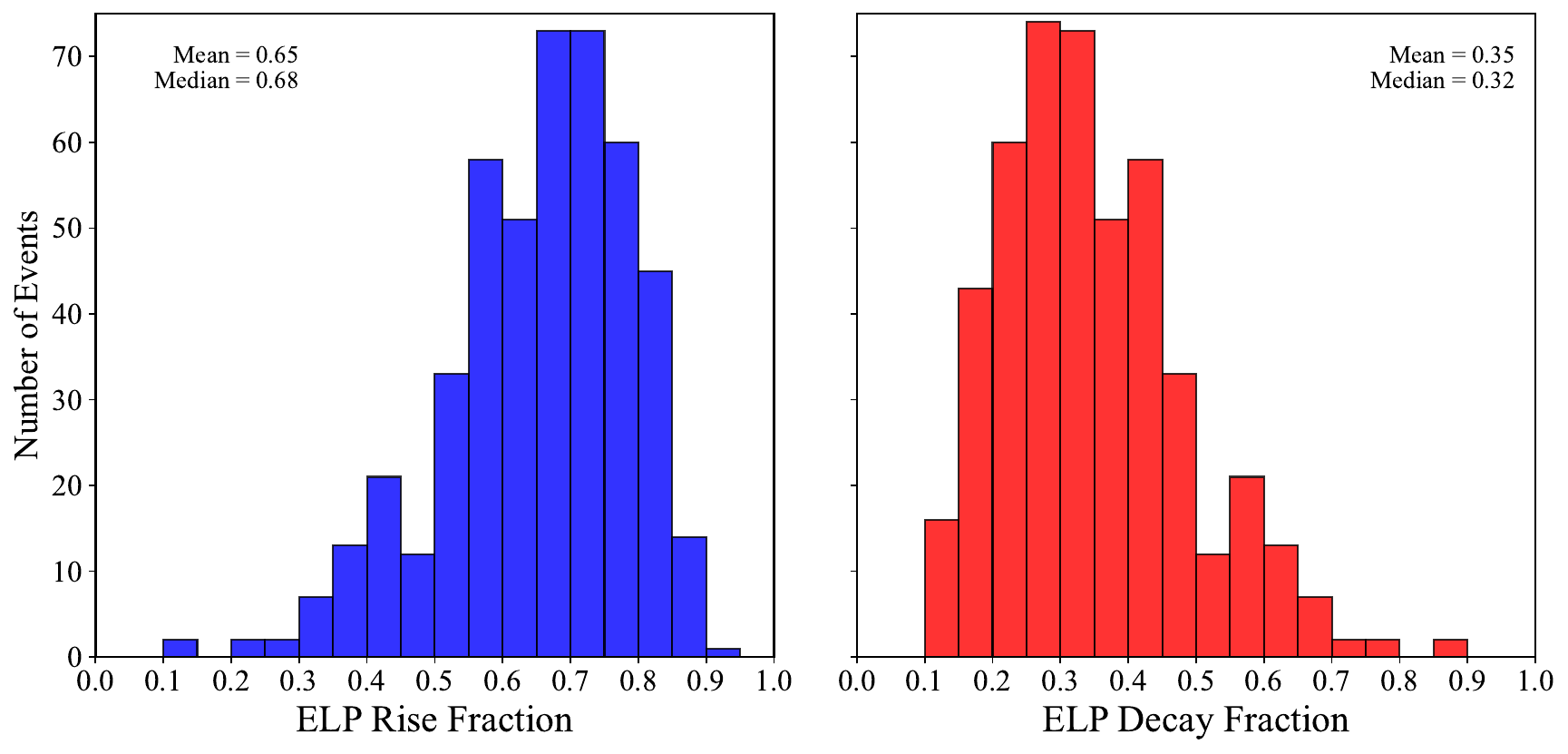}
\caption{Distributions of the Fe~{\sc{xvi}} late-phase rise fraction (left; blue) and decay fraction (right; red) for the 467 validated events. The histograms are constructed in bins of 0.05. Each fraction is defined relative to the total ELP duration ($\mathrm{{t_{rise}}/t_{ELP}}$ and $\mathrm{{t_{decay}}/t_{ELP}}$). Mean and median values are displayed in the upper-right corner of each panel.}
\label{fig:rise_decay_frac}
\end{figure}

To quantify the relative contributions of the two stages of the late-phase, the Fe~{\sc{xvi}} rise and decay times were expressed as fractions of the total late-phase duration. Figure~\ref{fig:rise_decay_frac} shows the relative fractions of the ELP attributed to the rise (left panel) and decay phases (right panel), respectively. Both fractions exhibit broad variability across events, but the late-phase is typically dominated by the rise phase. The blue bars, representing the rise fraction, are centred above 0.5 (ranging from approximately 0.6--0.8), while the red bars, representing the decay fraction, peak below 0.5 (around 0.2--0.4). Averaged over all events, the typical rise and decay fractions were found to be 65~\% and 35~\%, respectively. This behaviour is consistent with the skew distribution reported above and has implications for the energy-release and transport mechanisms that sustain the ELP. In particular, shorter rise times may indicate more rapid secondary energy injection, such as continued or delayed magnetic reconnection, whereas broader, decay-dominated profiles are consistent with cooling-dominated evolution in extended post-flare loop systems. These interpretations are explored further in Section~\ref{sec:discussion}.

\subsection{Rise and Decay Rates}\label{subsec:rates}

To further assess the change in ELP flux over time, the rates at which the Fe~{\sc{xvi}} emission increased and decayed during the ELP were examined. These rates were calculated as the flux change per unit time during the rise and decay intervals. For each event, the rise rate was defined as $\rm R_{rise} =(F_{\rm peak} - F_{\rm start})/(t_{\rm peak}-t_{\rm start})$ and the decay rate as $\rm R_{decay} =(F_{\rm peak} - F_{\rm end})/(t_{\rm end}-t_{\rm peak})$. Here, $F_{start,~ peak,~end}$ are the respective flux values at the the start, peak, and end times of the ELP emission ($t_{start,~peak,~end}$; Section~\ref{subsec:elp_detection}).

\begin{figure}[h!]
\centering
\includegraphics[width=.7\linewidth]{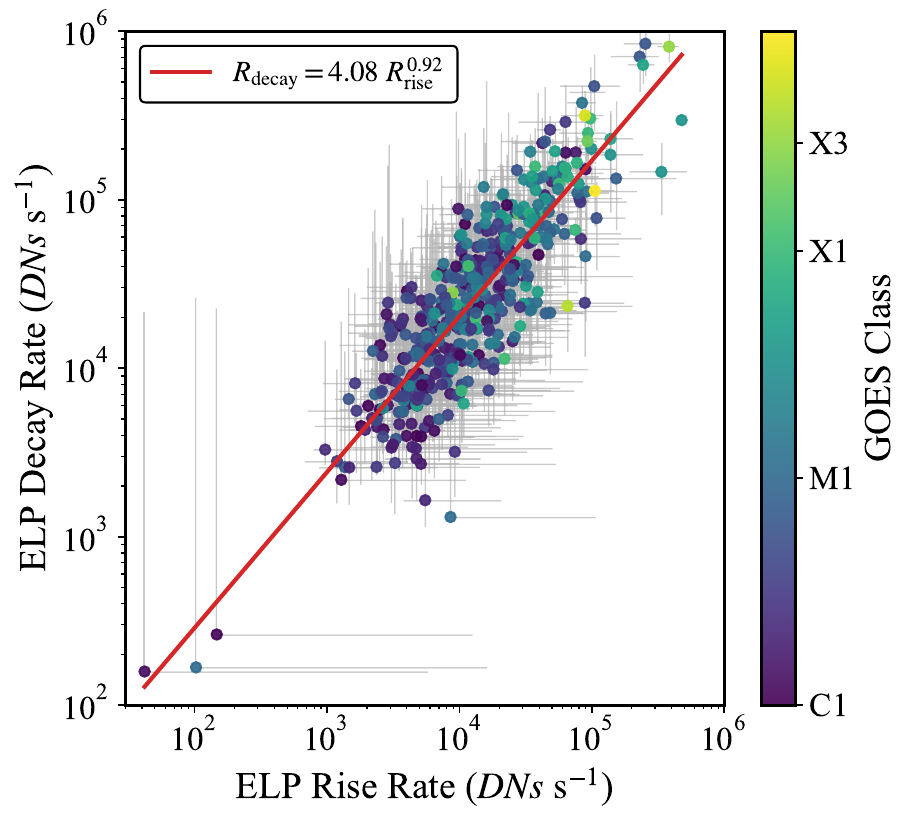}
\caption{Relationship between the rise and decay rates of the Fe~{\sc{xvi}} late-phase emission, derived for the 467 validated ELP events. Each point represents an individual flare, colour-coded by GOES class. Error bars denote the $1\sigma$ confidence intervals derived from 1000 Monte Carlo resamplings, where measured fluxes were perturbed by Gaussian noise assuming a 5~\% fractional photometric uncertainty. The red line indicates the best-fit power-law relation, $\rm R_{decay} = 4.08~R_{rise}^{0.92}$, obtained from a log-log linear regression. Pearson and Spearman correlation coefficients were $r = 0.79$ and $\rho = 0.76$, respectively.}
\label{fig:rise_decay_rate}
\end{figure}

Figure~\ref{fig:rise_decay_rate} presents the relationship between the rise and decay rates ($\rm R_{rise},~\rm R_{decay}$) across the sample. A clear correlation spanning more than three orders of magnitude was found, with only a small subset of low-rate events showing larger uncertainties due to reduced signal-to-noise. A log-log linear regression yielded a power-law relation of $\rm R_{decay} = 4.08~R_{rise}^{0.92}$, with a Pearson correlation coefficient of $r = 0.79 \pm 0.01$ and a Spearman rank coefficient of $\rho = 0.76 \pm 0.01$, both indicating a strong positive correlation. The quoted $1\sigma$ uncertainties were estimated from 1000 Monte Carlo resamplings of each event, in which the flare and ELP fluxes were perturbed by Gaussian noise adopting a 5~\% fractional photometric uncertainty, consistent with the short-term relative radiometric stability of AIA (\citealt{Boerner2012AIACalibration, Boerner2014AIACalibration}), and the rates were recomputed to propagate per-point measurement errors. A clear dependence on flare magnitude is also evident, with larger flares exhibiting systematically faster rise and decay rates. The near-unity slope demonstrates that these rates increase in a coherent manner, implying that the growth and decay of the Fe~{\sc{xvi}} late-phase emission may be linked processes.

\subsection{Impulsivity and Relative Flux} \label{subsec:impulsivity}

\begin{figure}[h!]
\centering
\includegraphics[width=.7\linewidth]{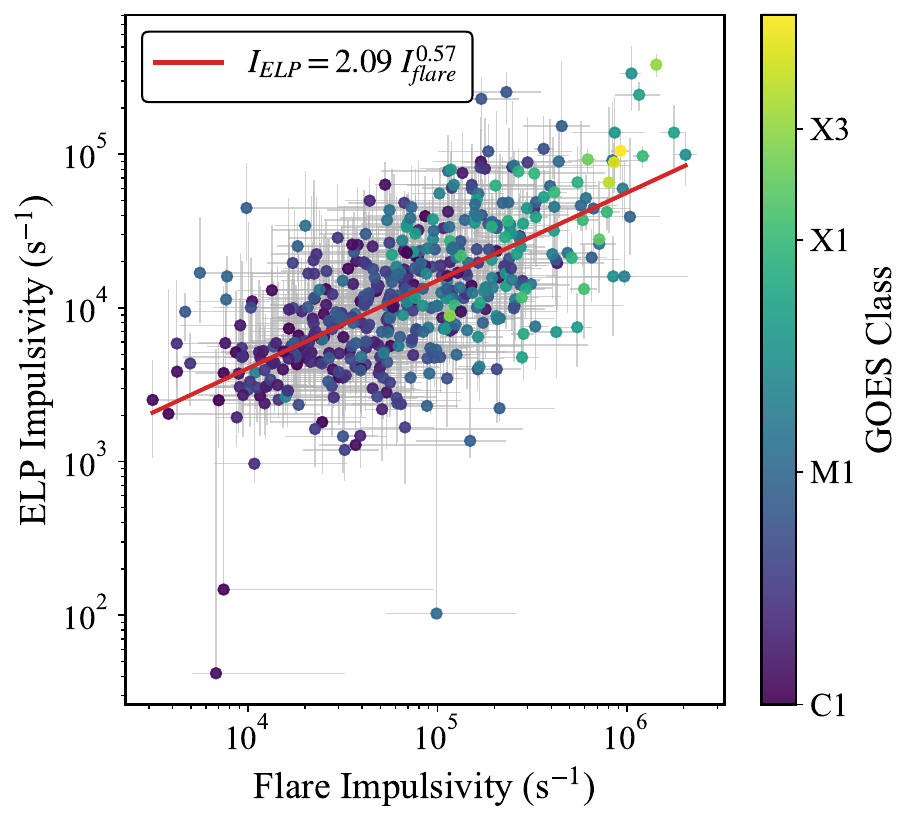}
\caption{Relationship between the impulsivity of the main flare and the corresponding ELP, calculated from Fe~{\sc{xvi}} emission for the 467 validated events. Each point represents an individual flare-ELP pair, coloured by GOES class. Error bars denote the $1\sigma$ uncertainty, including the propagated timing uncertainty and an additional uncertainty in the enhancement term corresponding to possible saturation. The red line shows the best-fit power-law relation obtained from a log-log linear regression, $\rm I_{ELP} = 2.09~I_{flare}^{0.57}$, with Pearson and Spearman correlation coefficients of $r = 0.60$ and $\rho = 0.61$, respectively.}
\label{fig:imulsivity}
\end{figure}

To examine the rate and relative strength of energy release during the ELP and how this compares to the associated initial flare, an adapted impulsivity parameter was calculated for the main flare and the corresponding ELP for each event in the sample, defined in Section~\ref{subsec:elp_metrics}. Figure~\ref{fig:imulsivity} presents the relationship between flare and ELP impulsivity for all validated events, with the points once again coloured by GOES class. A clear positive correlation is evident between the impulsivities of the two phases, spanning over three orders of magnitude in flux rate, albeit with a moderate degree of scatter. The best-fit power-law relation, obtained from a log-log linear regression, was $\rm I_{ELP} = 2.09~I_{flare}^{0.57}$, with a Pearson correlation coefficient of $r = 0.60 \pm 0.02$ and a Spearman rank coefficient of $\rho = 0.61 \pm 0.02$. Uncertainties were estimated using a similar Monte Carlo process as those used for the rise and decay rates, shown as grey error bars in Figure~\ref{fig:imulsivity}, with an additional uncertainty term included to account for possible saturation-related systematics inferred from the desaturation uncertainty analysis (see Section~\ref{subsec:elp_metrics}). A clear dependence on the flare magnitude is present, where the larger flares feature both a greater impulsivity in the initial flare and subsequent ELP. The sub-linear slope ($<1$) of the best-fit power law indicates that ELP impulsivity increases more slowly than flare impulsivity. These findings may have implications for the driving mechanism in the associated energy release of each phase (further discussed in Section~\ref{sec:discussion}).

\begin{figure}[h!]
\centering
\includegraphics[width=0.7\linewidth]{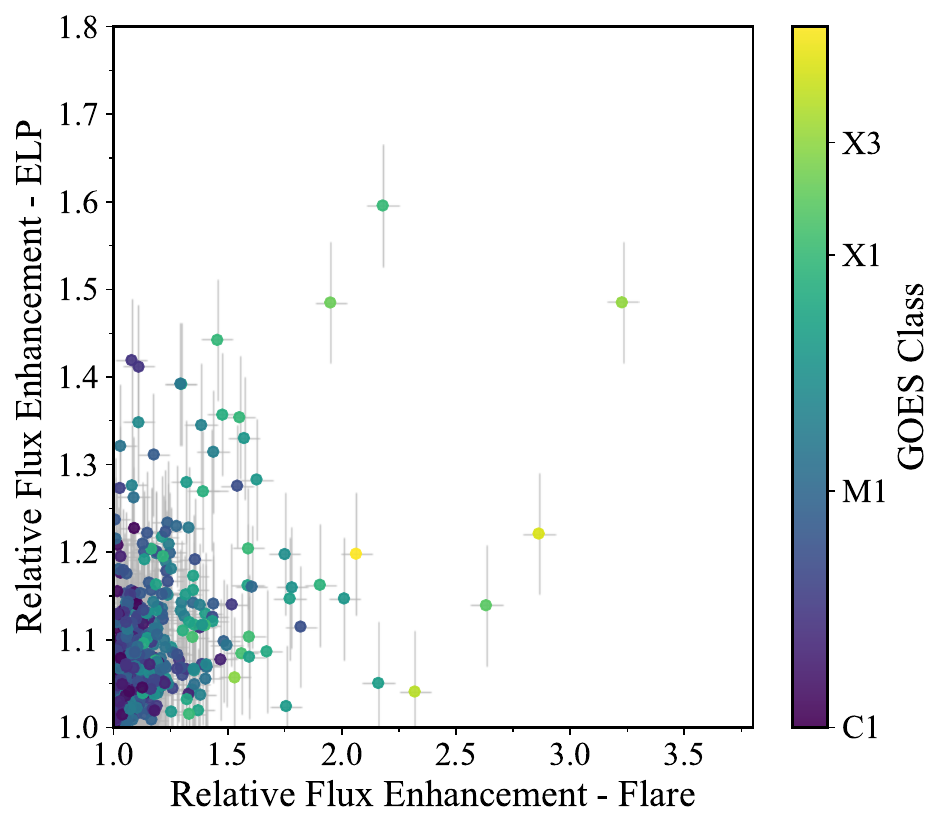}
\caption{Relationship between the peak relative flux enhancements of the Fe~{\sc{xvi}} emission during the main flare and the corresponding ELP for the 467 validated events. Each point represents an individual event, coloured by GOES class. The relative enhancement for each phase was calculated as the ratio of the peak flux to the fitted preflare background level. Error bars denote the $1\sigma$ uncertainties of 0.07 in the relative enhancement, based on the residual uncertainty associated with possible saturation effects inferred from the desaturation uncertainty analysis.}
\label{fig:enhance}
\end{figure}

To complement the impulsivity analysis, which incorporates temporal evolution, we also compared the relative enhancements of Fe~{\sc{xvi}} emission to isolate the magnitude of the flux increase independent of timescale (Figure~\ref{fig:enhance}). The mean flare enhancement was 1.17~$\pm$~0.23, while the mean ELP enhancement was 1.10~$\pm$~0.08~\%, corresponding to average flux increases of 17~\% and 10~\%, respectively. For individual events, we adopted a 1$\sigma$ uncertainty of 0.07 in the relative enhancement, based on the residual uncertainty associated with possible saturation effects inferred from the desaturation uncertainty analysis. The distribution shows no clear linear relationship between the two quantities, with the majority of events clustered toward lower enhancements and with all events sufficiently scattered. Several events displayed large flare enhancements without equivalent large increases in the ELP, thus no systematic dependence of ELP enhancement on flare class is apparent.  

Comparing the peak enhancements between the two phases, approximately 36~\% of the flares in the sample demonstrated Fe~{\sc{xvi}} enhancements that were greater in the ELP than in the main flare. Thus, only approximately a third of these events may be considered to be extreme ELP events \citep{Chen2020ELP}. It is important to note that due to the significant workload required to calibrate each individual event over such a large sample, all measurements of flux remain relative for this investigation. Further follow-up investigations may be conducted for a smaller sub-sample of flux-calibrated events to examine ELP energetics and absolute fluxes.

\subsection{Correlated Parameters}\label{correlated_params}

\begin{figure}[h!]
\centering
\includegraphics[width=1.\linewidth]{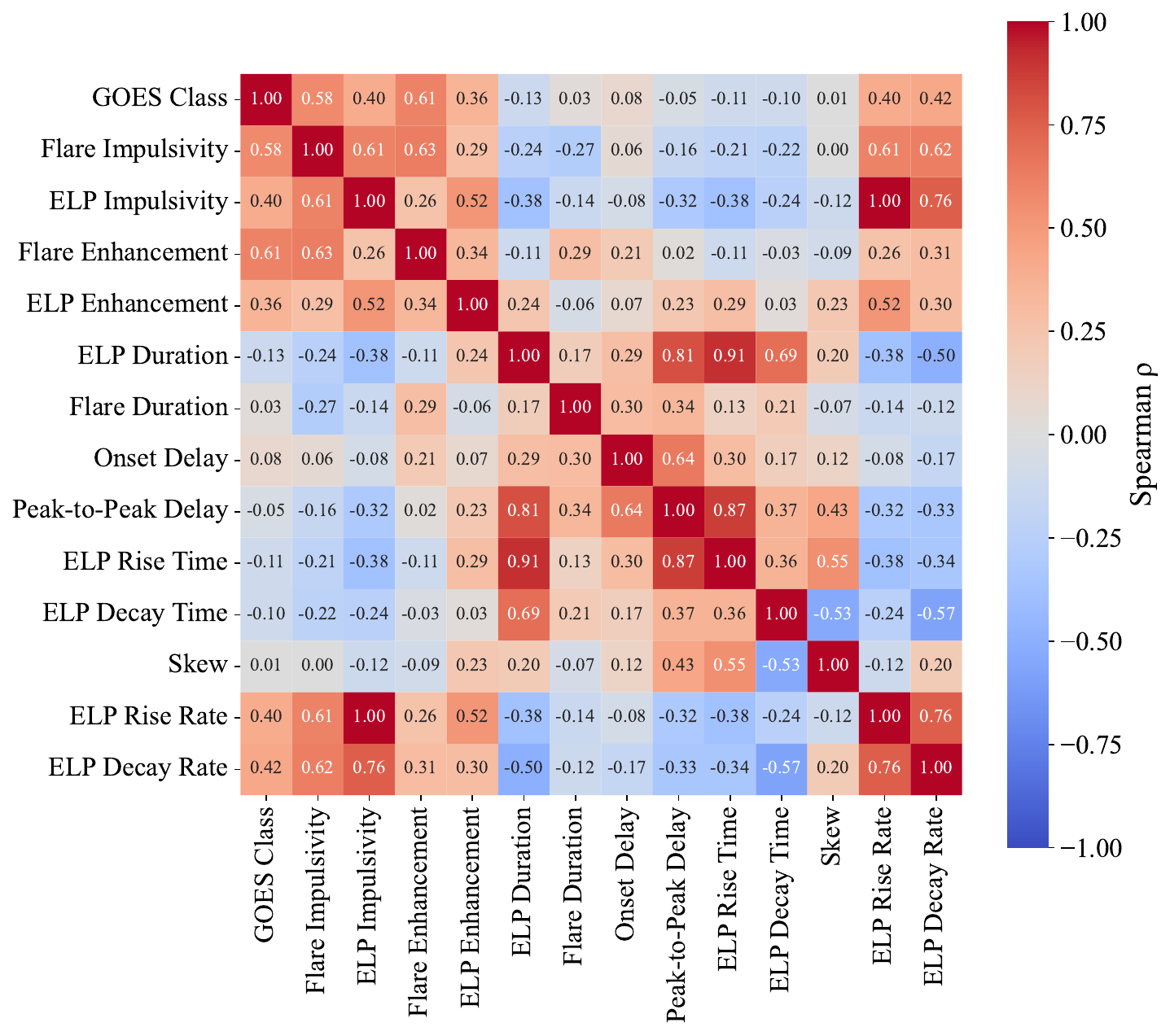}
\caption{Pairwise Spearman rank correlation matrix for the temporal and morphological parameters of the 467 validated ELP events. Each cell represents the correlation coefficient ($\rho$) between two quantities derived from the Fe~{\sc{xvi}} emission. The colour scale indicates the strength and sign of the correlation, with positive values in red and negative values in blue. The annotated values represent the mean $\rho$ obtained from 1000 bootstrap resamplings.}
\label{fig:corr_matrix}
\end{figure}

To explore interdependencies among the derived ELP parameters and test for previously unidentified relationships, pairwise Spearman correlation coefficients were computed between the parameters for all validated ELP events. This analysis considered the variables outlined in Section~\ref{subsec:temporal_elp} and \ref{subsec:impulsivity} for the flare and ELP emission. The Spearman rank correlation coefficient was adopted as the primary metric because it quantifies monotonic relationships without assuming linearity and is less sensitive to outliers than the Pearson correlation coefficient. Figure~\ref{fig:corr_matrix} presents the resulting pairwise Spearman correlation matrices.

Aside from the correlations already identified, several of the strongest correlations in Figure~\ref{fig:corr_matrix} arise directly from definitional coupling between the parameters, such as the relationship between ELP duration, rise time, and peak-to-peak delay. This is especially true for parameters such as impulsivity, enhancements, and rise and decay rates, as these are effectively separated only by scaling factors. These correlations therefore do not carry physical significance. This analysis does confirm and quantify the finding that several of those parameters have some dependence on the GOES class, and therefore the magnitude of the flare. However, the more informative relationships are those comparing flare parameters with ELP parameters, particularly those related to temporal evolution, which show uniformly low Spearman coefficients ($\rho < 0.30$). In particular, the flare duration is only weakly related to the ELP duration ($\rho = 0.17 \pm 0.04$) and to the ELP rise and decay times ($\rho = 0.13$ and $\rho = 0.21$, respectively). This lack of correlation between these metrics demonstrates that the temporal evolution of the late-phase is largely separate from that of the main flare despite their shared magnetic environment (this is further discussed in Section~\ref{sec:discussion}).

\subsubsection{Principal Component Analysis}\label{:subsubsec:pca}


To explore whether the flare and ELP parameters reflect a smaller number of underlying physical drivers, a Principal Component Analysis (PCA) was performed. PCA is a dimensionality-reduction technique that identifies orthogonal combinations of variables (principal components; PCs) that capture the largest fraction of variance in a multi-parameter dataset. Mathematically, PCA diagonalises the correlation matrix of the standardised parameters, yielding eigenvectors that define new orthogonal axes in parameter space and eigenvalues that quantify the variance captured along each axis. Each principal component is therefore a linear combination of the original variables, with coefficients referred to as loadings that indicate how strongly each parameter contributes to that axis and whether the relationship is positive or negative. The score of each event along a given component is obtained by projecting its standardised parameters onto the corresponding eigenvector. Rather than examining numerous pairwise correlations independently, PCA therefore tests whether the distribution of flare-ELP events in 13-dimensional parameter space can be described by a smaller number of dominant directions of variability. If a small number of components are found to account for a large fraction of the total variance, this would suggest that the flare–ELP system is governed by a limited set of dominant physical processes. Conversely, a more distributed variance structure would indicate intrinsically multi-dimensional behaviour.

\begin{figure}
\centerline{
    \hspace*{0.015\textwidth}
    \includegraphics[width=0.515\textwidth,clip=]{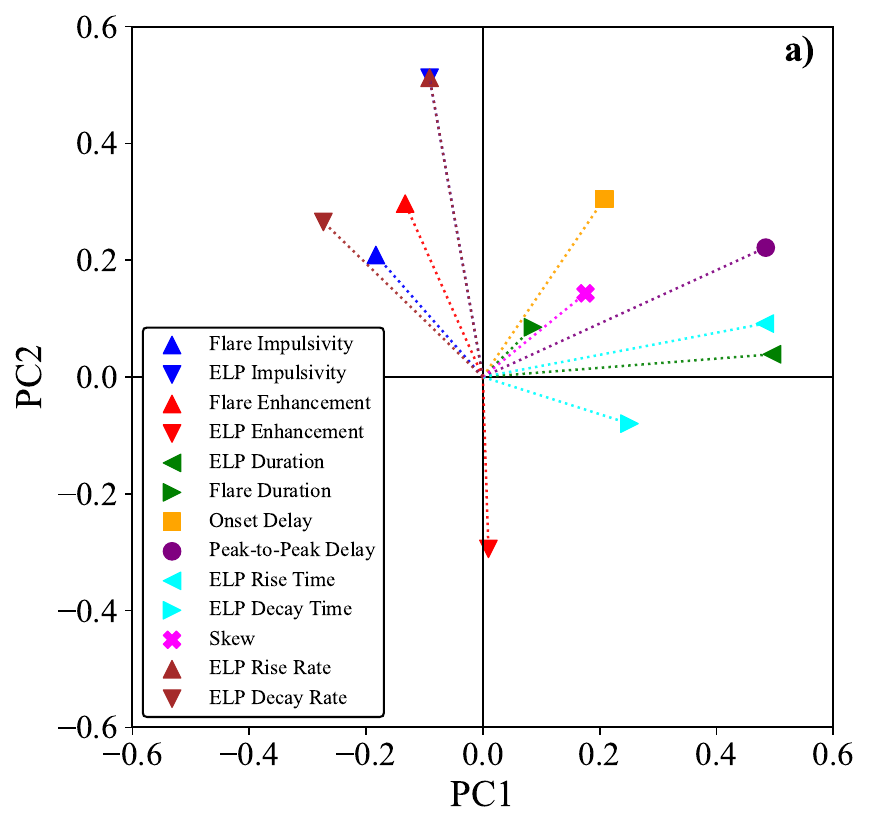}
    \hspace*{-0.03\textwidth}
    \includegraphics[width=0.515\textwidth,clip=]{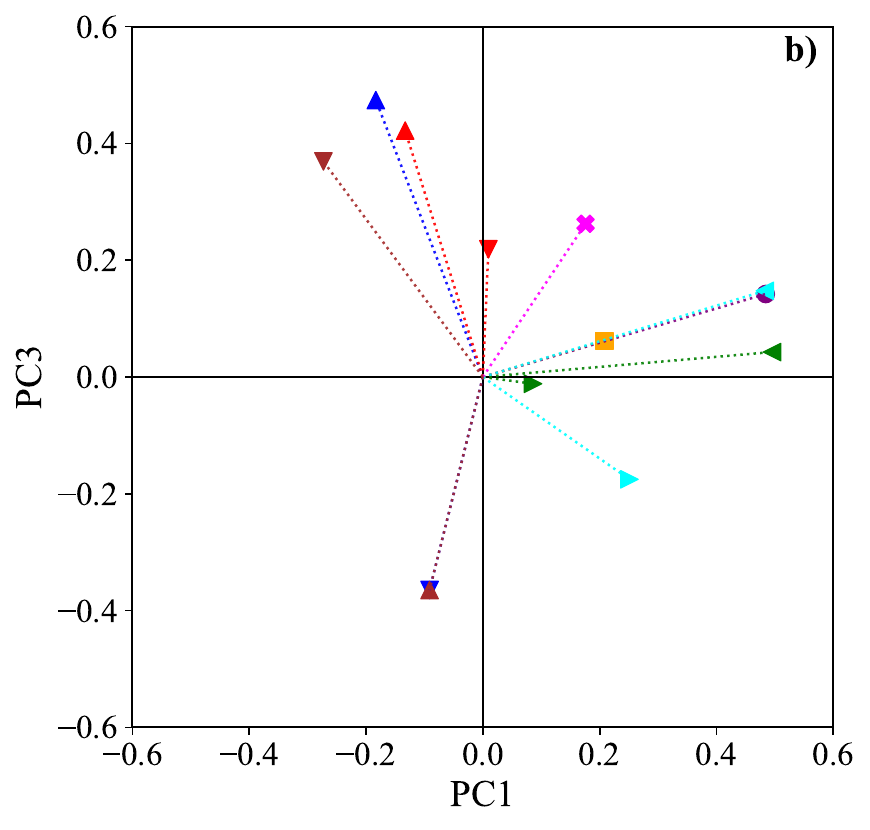}
}

\vspace{-0.32\textwidth}
\centerline{
    \Large \bf
    \hspace{0.00\textwidth} \color{black}{}
    \hspace{0.415\textwidth} \color{black}{}
    \hfill
}
\vspace{0.30\textwidth}

\caption{
PCA of flare and EUV late-phase parameters. Panel (a): two-dimensional projection of the parameter loadings in PCA space for PC1 and PC2. Panel (b): two-dimensional projection of the parameter loadings in PCA space for PC1 and PC3. Each marker represents an individual physical parameter derived from the time-series analysis in Section~\ref{sec:results}, and its position indicates how strongly that parameter contributes to the corresponding principal components.
}
\label{fig:PCA}
\end{figure}

The results of this analysis are presented in Figure~\ref{fig:PCA} and the loadings across each PC are summarised in Table~\ref{tab:pca_combined} along with the variances. Three principal components were found to capture up to approximately 60~\% of the total variance in the dataset, with PC1 representing 26.5~\%, PC2 17.8~\%, and PC3 15.3~\%. 

The loading projections shown in Figure~\ref{fig:PCA} illustrate how the original parameters align relative to one another within the reduced PCA space and identify which variables contribute most strongly to each principal component. In the PC1--PC2 plane (panel a), temporal measures of the ELP duration (0.50), peak-to-peak delay (0.48), and ELP rise time (0.48) extend predominantly along PC1, indicating that this axis captures the overall temporal scale of the late-phase evolution. Events with higher PC1 scores therefore correspond to longer-duration, more delayed late-phases, while lower scores reflect more temporally compact behaviour. In contrast, impulsivity-related parameters, particularly ELP impulsivity and ELP rise rate (both 0.51), project strongly along PC2, with secondary contributions from flare enhancement (0.30) and onset delay (0.30). PC2 therefore represents the rapidity or intensity of energy delivery into the late-phase system, distinguishing sharply rising late-phases from more gradual brightenings. The clear separation between temporal parameters (PC1) and rate-based parameters (PC2) indicates that the timescale of the late-phase and its heating rate constitute largely independent axes of variability.

The PC1--PC3 projection (panel b) further clarifies the structure of the parameter space. PC3 is characterised by strong positive loadings from flare impulsivity (0.47) and flare enhancement (0.42), alongside weaker or negative contributions from their ELP counterparts. This component therefore reflects the relative dominance of flare and late-phase intensity measures, differentiating events in which the flare exhibits comparatively stronger dynamical signatures from those in which the late-phase plays a proportionally larger role. The distribution of loadings across both projections reinforces that no single parameter defines the PCA structure; rather, multiple semi-independent contributions shape the dataset. While the loadings rarely exceed 0.5 in magnitude, the strongest contributors to each principal component fall within the range typically considered physically meaningful ($|$loading$|$ $\ge$ 0.45; \citealt{Jolliffe2011PCA}). The relationships between some high-loading parameters may reflect definitional overlap, and their physical interpretations remain speculative. Further discussion on the implications of this analysis is presented in Section~\ref{sec:discussion}.

\begin{table}[h!]
\footnotesize
\centering
\caption{
PCA results for the 13 flare and EUV late-phase parameters used in this study ($N=467$). (a) Loadings of each parameter on the first three principal components. (b) Explained variance ratios for PC1--PC3.
}
\label{tab:pca_combined}
\begin{tabular}{lccc}
\hline
\hline
\multicolumn{4}{c}{\textbf{(a) PCA Loadings}} \\
\hline
\multicolumn{1}{c}{Parameter} &
\multicolumn{1}{c}{PC1} &
\multicolumn{1}{c}{PC2} &
\multicolumn{1}{c}{PC3} \\
\hline
Onset Delay           & $0.208$  & $0.305$  & $0.061$ \\
Peak-to-Peak Delay    & $0.484$  & $0.222$  & $0.142$ \\
Flare Duration        & $0.086$  & $0.085$  & $-0.012$ \\
ELP Duration          & $0.495$  & $0.039$  & $0.043$ \\
ELP Rise Time         & $0.483$  & $0.092$  & $0.148$ \\
ELP Decay Time        & $0.251$  & $-0.080$ & $-0.175$ \\
Skew                  & $0.176$  & $0.143$  & $0.262$ \\
ELP Rise Rate         & $-0.091$ & $0.512$  & $-0.365$ \\
ELP Decay Rate        & $-0.273$ & $0.265$  & $0.369$ \\
Flare Impulsivity     & $-0.183$ & $0.209$  & $0.474$ \\
ELP Impulsivity       & $-0.091$ & $0.512$  & $-0.365$ \\
Flare Enhancement     & $-0.133$ & $0.297$  & $0.422$ \\
ELP Enhancement       & $0.009$  & $-0.294$ & $0.219$ \\
\hline
\multicolumn{4}{c}{} \\
\multicolumn{4}{c}{\textbf{(b) Explained Variance Ratio}} \\
\hline
\multicolumn{1}{c}{PC1} &
\multicolumn{1}{c}{PC2} &
\multicolumn{1}{c}{PC3} &
\multicolumn{1}{c}{} \\
\hline
0.265 & 0.178 & 0.153 & \\
\hline
\end{tabular}
\end{table}

\section{Discussion}\label{sec:discussion}

The results presented in this study provide new insight into the occurrence and temporal behaviour of the EUV late-phase in solar flares, based exclusively on Fe~{\sc{xvi}} emission observed by SDO/AIA. Across the 15~year dataset, validated ELPs were found in approximately 9~\% of isolated flares, which is lower than previously reported in studies using disk-integrated SDO/EVE observations \citep{Woods2011ELP, Woods2014EUVLatePhase, Chen2020ELP, Ornig2025ELPStats}. This reduction is not attributable to a single factor, but likely reflects a combination of differences in methodology, instrumentation, and sample composition. For example, the present work incorporates a more temporally expansive and statistically representative flare sample that includes both solar maximum and minimum conditions, and extends the flare selection threshold down to GOES C1.0, compared to C3.0 or higher in prior studies. Both of these choices can reduce the relative ELP detection rate by including a larger proportion of weaker or morphologically simple events, which are less likely to exhibit a pronounced late-phase.

The exclusive use of spatially resolved AIA observations, rather than full-disk EVE irradiance data, also contributes to statistical differences. A direct comparison between instruments revealed a 20~\% classification disagreement, highlighting the role of instrumental design and observational context. Several factors can lead to mismatches, including differing sensitivity to Fe~{\sc{xvi}}, distinct bandpasses, and systematic artefacts \citep{Greatorex2024Instruments}. In the majority of discrepant cases, AIA detects a clear ELP while EVE does not. This often occurs during periods of high solar activity, when emission from multiple active regions contributes to the full-disk EVE signal. In such scenarios, contamination from unrelated regions can obscure or dilute the flare-associated Fe~{\sc{xvi}} enhancement, particularly for weaker or spatially compact events. The spatial resolution of AIA circumvents this by isolating the flare source region, allowing robust identification even when the integrated signal remains ambiguous. Conversely, a smaller fraction of cases exhibit the opposite behaviour, whereby an ELP is identifiable in EVE but not in AIA. These typically involve events where the late-phase emission extends beyond the 500~$\times$~500~arcsec AIA cutout or is spatially fragmented across disconnected regions. In such cases, the fixed cutout misses the full extent of the emission, while EVE retains sensitivity to the global EUV response. Additional complications may arise from coronal dimming, post-eruptive absorption, or magnetic reconfiguration, all of which can redistribute or obscure ELP emission outside the AIA field of view \citep{Lopez2017Dimming, Dissauer2018Dimming, Toriumi2019ActiveRegions}. Ultimately, this study was necessarily focussed on spatially constrained observations, which while not immune to observational shortcomings, gave a robust, isolated detection of the ELP from flare active regions. The lower overall occurrence rate reported here should therefore be understood as the result of a more inclusive sample, broader temporal coverage, and spatially resolved methodology, rather than a limitation or contradiction of earlier work.

In this study, we identify that only around one third of the ELP events in our sample qualify as extreme late-phases, where the Fe~{\sc{xvi}} flux exceeds that of the main flare. This is significantly fewer than reported by \citet{Chen2020ELP} and \citet{Ornig2025ELPStats}, and likely arises from our inclusion of weaker flares, the application of a more permissive ELP threshold, and the use of a much larger sample size. No significant dependence of ELP occurrence on the solar cycle was identified, with the detection rate remaining broadly consistent throughout the 15~year period, which is the longest temporal baseline examined to date. This consistency suggests that while the absolute number of ELPs scales with the overall flare rate, their relative frequency remains constant across changing activity levels. It was found that ELP occurrence has only a slight dependence on flare magnitude, consistent with the implications of an independence from solar cycle. A partial tendency toward the mid M-class range was found, which may reflect the greater magnetic energy reservoir and more extensive post-eruptive loop systems typically associated with larger events, increasing the likelihood of producing a detectable secondary enhancement. However, it is important to recognise that the operational definition adopted here requires a secondary Fe~{\sc{xvi}} enhancement, which traces plasma at approximately 2.5~MK. This requirement may preferentially select events in which post-flare loop systems are heated to, or beyond, warm coronal temperatures, potentially contributing in part to the observed class tendency. The distribution therefore likely reflects a combination of physical scaling and observational definition. However, smaller flares may still exhibit measurable and, in some cases, extreme late-phase enhancements in Fe~{\sc{xvi}}, indicating that the phenomenon is not restricted solely to the most energetic events \citep{Liu2015ExtremeELP, Bekker2024ELP}.

Beyond occurrence statistics, the temporal morphology of the Fe~{\sc{xvi}} emission provides further insight into the nature of the late-phase. A strong positive correlation was identified between the rise and decay rate of the ELP ($r = 0.79$, $\rho = 0.76$). Indicating that flares with a faster rate of increase also tended to have a faster decay rate. A clear positive correlation was also found between the impulsivities of the flare and ELP phases ($r = 0.60$, $\rho = 0.61$), indicating that, on an event-by-event basis, more impulsive flares tend to be followed by proportionally more impulsive late-phases. These relationships may hint at a coupling between the energy release of the two phases, rather than a single common ELP mechanism operating identically in all cases.

An important result from this study is the strong correlation between the Fe~{\sc{xvi}} rise and decay rates during the late-phase. The rates span over three orders of magnitude yet follow a coherent power-law relation, with greater rise rates accompanied by proportionally greater decay rates. This behaviour suggests that the processes governing the increase and subsequent decline of the warm coronal emission evolve coherently. If the late-phase is powered by continued or secondary reconnection, then both the brightening and cooling rates may scale with the magnetic energy release and the geometry of the overlying loop system \citep{Woods2011ELP, Dai2013ELP_Reconnection, Liu2015ExtremeELP}. Conversely, in cooling-dominated cases, long and dense coronal loops naturally produce slower rises and slower decays, as both conductive and radiative cooling timescales increase with loop length and plasma density \citep{Cargill1995Coolingplasma, Woods2014EUVLatePhase}. The near-unity slope of the relation therefore accommodates both scenarios and implies that, irrespective of whether renewed heating or extended cooling dominates, the temporal evolution of the late-phase is governed by a common underlying physical structure in the active region. The small population of low-rate events with large uncertainties likely reflects cases where the late-phase emission is weak relative to background levels or where the geometry produces very gradual evolution \citep{Woods2011ELP}.

Another key result from this study is the scaling relationship between the impulsivity of the main flare and that of the ELP. One possible conclusion from this finding is that both phases share a common dependence on the magnetic properties of their associated active region. \citet{Tamburri2024Impulsivity} reported that larger EUV impulsivities correspond to stronger magnetic reconnection rates, implying that the impulsive evolution is magnetically mediated. Although we do not directly examine the magnetic topology of the ELP regions, other studies have similarly found links between magnetic configuration and late-phase emission \citep{Woods2011ELP, Ornig2025ELPStats}. Despite this, it must be acknowledged that two potential mechanisms can produce late-phase emission: continued or secondary reconnection and extended cooling of large, overlying flare loops. Variability in impulsivity among events may therefore arise from differences in the dominant process. In cooling-dominated cases, the rise rate of Fe~{\sc{xvi}} emission is governed primarily by conductive and radiative cooling timescales, which depend on loop length, temperature, and density \citep{Cargill1995PlasmaCooling}. This naturally broadens the impulsivity distribution and may explain the scatter around the observed scaling relation. Some of the more weakly correlated events could correspond to confined flares that do not significantly alter the overlying magnetic structure \citep{Aschwanden2009Confined, Woods2011ELP}, where extended cooling dominates over renewed energy release. Ultimately, it is not possible to determine the dominant driver of each ELP solely from the emission timeseries. Nevertheless, the proportional relationship between flare and ELP impulsivities carries important implications for the terrestrial response, whereby rapid energy injection into the upper atmosphere during the main flare may often be followed by a proportionally rapid secondary injection during the late-phase, on timescales characterised by the onset and peak-to-peak delays. We also find that greater magnitude flares typically exhibit a more impulsive ELP but not necessarily a higher ELP occurrence rate, indicating that additional factors, such as the active region topology, remain decisive in determining whether a late-phase forms at all.

The PCA results further support the interpretation that ELP evolution is governed by several physically distinct processes. A summary of the PCA results is presented for succinctness in Table~\ref{tab:PCA_summary}. The modest variance captured by the leading component (26.5\%), and the distributed nature of the loadings across all three components, indicate that no single parameter or simple scaling relation governs ELP behaviour. Instead, the system is governed by partially independent factors that contribute to the observed variability. Within this framework, PC1 is most naturally interpreted as a temporal scale axis, reflecting the covariance between ELP duration, rise time, and peak-to-peak delay. This alignment suggests that the characteristic timescale of the late-phase evolution varies independently of other dynamical measures. Such temporal variability is broadly consistent with scenarios in which loop geometry, magnetic connectivity, and reconnection timescales influence the evolution of post-flare arcades \citep{Woods2011ELP, Chen2020ELP}. However, the relatively limited variance explained by PC1 implies that temporal scaling alone cannot account for the diversity of ELP behaviour. PC2 is primarily associated with ELP impulsivity and rise rate, with secondary contributions from flare enhancement and onset delay. This component therefore reflects variability in the rate or intensity of energy delivery into the late-phase system. The inclusion of flare-related quantities on this axis is consistent with the previously identified correlations between flare and ELP impulsivity, and may indicate that more dynamically evolving active regions tend to produce both more impulsive flares and more rapidly brightening late-phases. This interpretation is compatible with reconnection-rate-dependent impulsivity scenarios discussed in the literature \citep{Tamburri2024Impulsivity}, although PCA alone does not establish a causal mechanism. PC3 differentiates flare-dominated intensity signatures from those more strongly associated with the late-phase, as indicated by the opposing loadings of flare and ELP impulsivity-related parameters. Rather than demonstrating strict energy partitioning, this component more conservatively reflects variability in the relative prominence of flare and late-phase dynamical measures within individual events. Importantly, the distributed variance and moderate loading magnitudes reinforce the conclusion that ELP evolution is not reducible to a small set of governing parameters. Instead, temporal scale, impulsive heating characteristics, and flare-ELP contrast appear to vary semi-independently across the population. In this sense, the late-phase exhibits identifiable trends, but is important to note that these should not be over interpreted. Future work incorporating magnetic field diagnostics and energetics modelling will be required to determine a decisive driving mechanism of late-phase emission.

\begin{table}[h!]
\centering
\caption{Summary of the three principal components derived from PCA of flare and ELP parameters. Each component is labelled with its physical interpretation, the percentage of variance it explains, and the three primary contributing parameters with their corresponding loading values.}
\begin{tabular}{clcl}
\hline
\textbf{PC} & \textbf{Physical Interpretation} & \textbf{Explained Variance (\%)} & \textbf{Primary Parameters} \\
\hline
PC1 & ELP Temporal Evolution & 26.5 & 
\begin{tabular}[c]{@{}l@{}}ELP Duration (0.50) \\ ELP Rise Time (0.48) \\ Peak-to-Peak Delay (0.48)\end{tabular} \\
PC2 & Energy Release & 17.8 & 
\begin{tabular}[c]{@{}l@{}}ELP Impulsivity \& Rise Rate (0.51) \\ Flare Enhancement (0.30) \\ Onset Delay (0.30)\end{tabular} \\
PC3 & Flare-ELP Energy Partition & 15.3 & 
\begin{tabular}[c]{@{}l@{}}Flare Impulsivity (0.47) \\ Flare Enhancement (0.42) \\ ELP Decay Rate (0.37)\end{tabular} \\
\hline
\end{tabular}
\label{tab:PCA_summary}
\end{table}

The onset of the ELP typically followed the end of the main flare by approximately 19~minutes, while the delay between the flare and ELP peaks was around 88~minutes, in good agreement with previous studies \citep{Woods2011ELP, Woods2014EUVLatePhase, Ornig2025ELPStats}. Slight discrepancies in timing may arise from the use of 60~s cadence SDO/AIA images compared to 10~s SDO/EVE data and the larger number of events analysed, but these differences are minor. Only 36~\% of the sample exhibited onset delays greater than 20~minutes, beyond which the distribution became significantly more dispersed. This suggests that longer delays may be governed by the cooling timescales of the overlying flare loops, which are more variable than the reconnection timescales likely responsible for the faster onsets

The average ELP duration was approximately 93~minutes, consistent with previous observations using SDO/EVE. This duration was found to be unrelated to either the flare duration or its magnitude. As reported by \citet{Ornig2025ELPStats}, the ELP duration is well correlated with the peak-to-peak delay, here with a comparable Spearman coefficient ($\rho = 0.81$). We also find that both the ELP duration and the peak-to-peak delay are strongly correlated with the ELP rise time ($\rho = 0.91$ and $\rho = 0.87$, respectively). This suggests that the rise phase dominates the overall ELP evolution, consistent with the results in Section~\ref{subsubsec:temporal_morph}, where the rise phase was typically found to last twice as long as the decay phase and to constitute roughly two thirds of the total ELP duration. Interestingly, the temporal parameters defining the ELP show minimal correlation with those of the main flare (typically $|\rho| < 0.3$), indicating that while the two phases are temporally distinct, their impulsivity relationship implies that they remain linked by their driving mechanisms.

Together, these relationships provide key context for understanding the physical mechanisms underlying ELP emission and serve as a benchmark for future classification or predictive models. Moreover, the occurrence statistics, solar cycle invariance, and measured temporal parameters presented here are directly applicable to assessing the terrestrial impact of ELP emission, which is invaluable to our understanding of the Sun-Earth connection. We note, however, that the present analysis relies on relative Fe~{\sc{xvi}} fluxes rather than absolute irradiances and does not directly incorporate magnetic field measurements. Future work combining these temporal diagnostics with vector magnetograms and irradiance calibration will be critical for distinguishing reconnection-dominated and cooling-dominated late-phases and quantifying their respective energy contributions.

Furthermore, \cite{Woods2011ELP} suggested that EUV late-phase emission may be associated with giant arch systems, which are large, high-lying post-eruptive loop structures that can persist and grow for many hours. However, the disk-integrated nature of their analysis limited detailed morphological assessment of these structures. In this study, the validation procedure requires that the late-phase emission originate from a spatially distinct loop system, and in some cases these appear as overlying arcades broadly consistent with post-eruptive loop systems discussed in giant arch analyses (\citealt{Jager1985GiantArches, Svetska1984GiantArches, Svestka1995GiantArches, Svestka1996GiantArches, Svestka1997GiantArches, Svestka1998GiantArches, West2015GiantArches}). Nevertheless, the morphology across the 467 validated events is highly variable, and no quantitative classification of loop height, extent, or long-term evolution was performed. Moreover, the typical late-phase durations reported here  are substantially shorter than the extreme multi-hour or day-long timescales associated with some giant arch events. While a subset of ELPs may represent shorter-lived analogues of such systems, the two phenomena cannot be considered equivalent on the basis of the present statistical analysis. A dedicated morphological investigation would be required to determine whether a distinct subpopulation of giant arch ELPs exists within the broader sample.

While this study focuses on Fe~{\sc{xvi}} emission, the Fe~{\sc{xv}} (284~\AA) line is known to exhibit a similarly distinct late-phase component and has been shown to have a stronger influence on ionospheric ionisation \citep{Bekker2024ELP}. Given their closely overlapping formation temperatures (2.5--4.0~MK) and correlated behaviour during flares, Fe~{\sc{xvi}} serves as a physically justified proxy for Fe~{\sc{xv}} in characterising the late-phase evolution. Accordingly, the diagnostics presented here are expected to be directly applicable to studies of ionospheric response to ELP emission.

\section{Conclusions}\label{Conclusions}\label{sec:conclusions}

This study presents the statistical analysis of 467 ELP events identified from a sample of 5335 isolated flares observed by SDO/AIA between 2010 and 2025, constituting the largest and most temporally extended analysis of EUV late-phase behaviour to date. This work focuses on the occurrence rate, class dependence, and solar cycle variability of the ELP, as well as on several metrics that have implications for understanding the physical origin of ELP emission and its corresponding impact on the terrestrial atmosphere.

The overall ELP occurrence rate was approximately 9~\%, lower than in previous analyses based on disk-integrated SDO/EVE data. This difference may be attributed to both the larger, more representative flare sample, and the exclusive use of SDO/AIA measurements. No significant dependence of ELP occurrence on the solar cycle was identified, and only a weak dependence on flare magnitude was observed, with a modest preference for the low-mid M-class range (M3.0--M5.9). The rise and decay rates of the ELP were found to be tightly correlated over three orders of magnitude, suggesting that the processes governing the growth and decline of warm coronal emission evolve coherently. Additionally, a clear scaling relation was found between the impulsivity of the main flare and that of the ELP, indicating a potential link between the two phases through their shared magnetic environment. The typical onset and peak-to-peak delays of 19 and 88~minutes, respectively, together with the average ELP duration of 93~minutes, are consistent with earlier studies but derived here from a more extensive dataset. A pairwise correlation analysis of the temporal parameters of both phases revealed minimal dependence of the ELP evolution on the temporal characteristics of the flare, reinforcing the notion that, while energetically linked, the ELP is temporally distinct from the main flare. Principal Component Analysis provided a complementary view of the global covariance structure of the flare–ELP parameter space. The leading components collectively captured moderate fractions of the total variance, with PC1 primarily associated with late-phase temporal scale, PC2 with impulsive heating characteristics, and PC3 with the relative prominence of flare and late-phase intensity measures. However, no single axis dominated the dataset, reinforcing that ELP behaviour cannot be reduced to a simple scaling relation or single controlling parameter. Instead, the late-phase appears to reflect the interplay of multiple semi-independent processes. These interpretations remain descriptive, but provide a structured basis for future investigations incorporating magnetic and energetic diagnostics.


The results presented here provide a benchmark characterisation of the behaviour of the EUV late-phase, further highlighting that while the ELP is relatively rare, its occurrence and dynamics are likely governed by active-region magnetic structure and plasma-cooling processes, although these were not explicitly explored in this analysis. The statistical relationships reported here and the derived parameters are directly relevant to models of flare-associated emission such as the \textit{Flare Irradiance Spectral Model} (FISM; \citealt{Chamberlin2007FISM, Chamberlin2020FISM, Chamberlin2025FISM3}) and provide an empirical foundation for assessing the ionospheric response to delayed EUV radiation. The analysis is necessarily limited by the use of relative, rather than absolute, Fe~{\sc{xvi}} fluxes and by the absence of direct magnetic field measurements. Future work combining these diagnostics with magnetic extrapolations and irradiance-calibrated datasets will be essential for identifying the physical origins of late-phases and quantifying their respective energy budgets.

Finally, this work builds upon and extends previous statistical studies of the EUV late-phase by taking full advantage of the long-term, multi-wavelength coverage provided by SDO/AIA. The continuity of AIA observations has enabled the first systematic investigation of ELP behaviour over nearly two solar cycles using exclusively spatially-resolved diagnostics. These data continue to provide an essential foundation for advancing our understanding of solar flare evolution, the dynamics of the solar atmosphere, and the influence of solar activity on the terrestrial environment.

\section{Acknowledgements}

H.J.G, R.J.C, and R.O.M would like to thank the UK’s Science \& Technology Facilities Council (ST/X000923/1) for supporting this research. A.N.O, S.B, and R.O.M would like to thank the European Office of Aerospace Research and Development (FA8655-22-1-7044-P00001) for supporting this research. This work was also supported by the International Space Science Institute (ISSI) in Bern, through ISSI International Team project \#24-618. The authors would like to thank the anonymous reviewer for their helpful comments, which have greatly contributed to the overall improvement of this work. 


\section{Data Availability}

No new data was created as part of this study. SDO/AIA data may be accessed through the \textit{Joint Science Operations Centre} (JSOC; \url{http://jsoc.stanford.edu/}). SDO/EVE data may be accessed via the EVE data site or directly here: \url{https://lasp.colorado.edu/eve/data_access/index.html}. SXR data are publicly accessible from GOES-16 via \url{https://www.ncei.noaa.gov/products/goes-r-extreme-ultraviolet-xray-irradiance} and from GOES-14 and -15 via \url{https://www.ncei.noaa.gov/products/goes-1-15/space-weather-instruments}. All observation data in this study were analysed using sunpy (\url{https://sunpy.org/}: \citealt{SunPy2020}) and aiapy packages (\url{https://zenodo.org/records/17114673}: \citealt{Barnes2020AIAPy}).




%

%



%
%

%
%
%
%
%
%
%

%
%
\bibliographystyle{plainnat}
\bibliography{references}  
%
%
%
%

\end{document}